\documentclass[aps,prd,showpacs,preprintnumbers,superscriptaddress,preprint,groupedaddress]{revtex4}
\usepackage{amsmath}
\usepackage{color}
\usepackage{graphicx}
\usepackage{bm}
\usepackage{hyperref}
\allowdisplaybreaks[3]

\def \be {\begin{equation}}
\def \ee {\end{equation}}
\def \beqa {\begin{eqnarray}}
\def \eeqa {\end{eqnarray}}
\def \beal#1 {\begin{align}#1\end{align}}
\def \bes#1 {\begin{equation}\begin{split}#1\end{split}\end{equation}}
\def \nn {\notag\\}
\newcommand{\bealig}[2]{\begin{alignat}{#1}#2\end{alignat} }
\def\ds {\displaystyle}
\def\mat#1{\begin{matrix}#1 \end{matrix} }
\def\pmat#1{\begin{pmatrix}#1 \end{pmatrix} }
\def\bmat#1{\begin{bmatrix}#1 \end{bmatrix} }
\def\Bmat#1{\begin{Bmatrix}#1 \end{Bmatrix} }
\def\vmat#1{\begin{vmatrix}#1 \end{vmatrix} }
\def\Vmat#1{\begin{Vmatrix}#1 \end{Vmatrix} }
\usepackage[normalem]{ulem}  
\renewcommand\sout{\bgroup \color{red} \ULdepth=-.5ex \ULset}
\def\SA#1{\textcolor{red}{#1}}

\begin{document}

\preprint{YITP-21-98}
\preprint{RIKEN-iTHEMS-Report-21}
\title{Derivative expansion in the HAL QCD method for a separable potential}

\author{Sinya~Aoki}
\email{saoki@yukawa.kyoto-u.ac.jp }
\affiliation{Center for Gravitational Physics,
Yukawa Institute for Theoretical Physics, Kyoto University, Kyoto 606-8502, Japan}
\affiliation{Theoretical Research Division, Nishina Center, RIKEN, Saitama 351-0198, Japan}
\author{Koichi~Yazaki}
\email{koichiyzk@yahoo.co.jp}
\affiliation{Interdisciplinary Theoretical and Mathematical Sciences Program (iTHEMS), RIKEN Saitama 351-0198, Japan} 

\date{\today}

\begin{abstract}
We investigate how the derivative expansion in the HAL QCD method works to extract physical observables,
using a separable potential in quantum mechanics, which is solvable but highly non-local in the coordinate system. 
We consider three cases  for inputs to determine the HAL QCD potential in the derivative expansion,
(1) energy eigenfunctions (2) time-dependent wave functions as solutions to the time dependent   Schr\"odinger equation
with some boundary conditions  (3) time-dependent wave function made by a linear combination of finite number of  
eigenfunctions at low energy to mimic the finite volume effect. 
We have found that, for all three cases, the potentials  provide reasonable scattering phase shifts even at the leading order of the derivative expansion,  and they give more accurate results as the order of the expansion increases.
By comparing the above results with those from the formal derivative expansion for the separable potential,
we conclude that the derivative expansion is not a way to obtain the potential but a method to extract physical observables such as phase shifts and binding energies, and  that the scattering phase shifts from the derivative expansion in the HAL QCD method converge to the exact ones much faster than those from the formal derivative expansion of the separable potential.
\end{abstract}


\maketitle

\section{Introduction}
Nowadays not only simple quantities such as hadron masses and matrix elements but also more complicated quantities such as hadron interactions can be extracted in lattice QCD.   
Hadron interactions have been investigated in lattice QCD mainly by two method.
One is the finite volume method\cite{Luscher:1990ux}, the other is the HAL QCD potential method\cite{Ishii:2006ec,Aoki:2009ji,Aoki:2012tk}.
While both methods more or less utilize a fact that the Nambu-Bethe-Salpeter ((NBS) wave function encodes information of the S-matrix in QCD~\cite{Luscher:1990ux,Lin:2001ek,Aoki:2005uf,Ishizuka:2009bx,Ishii:2006ec, Aoki:2009ji, Aoki:2012tk,Carbonell:2016ekx,Aoki:2013cra,Gongyo:2018gou}, they have their own pros and cons, which are different from each other.
In particular, systematic errors of these methods are very different.
Systematic errors associated with the finite volume method are well understood, once finite volume spectra are precisely determined. 
On the other hand, 
the non-local potential in the HAL QCD method, which by definition correctly reproduces the scattering phase shift,   
needs in practice  to be approximated by the derivative expansion, whose systematic errors are difficult to quantify. 
Indeed, there was some misunderstanding on this point in literature. See some correspondences in \cite{Yamazaki:2017gjl,Aoki:2017yru,Yamazaki:2018qut}.

In this paper, we investigate how the derivative expansion of the potential works in the HAL QCD method, by applying it to a solvable model in quantum mechanics, whose potential has a separable form.
Separable potentials are  suitable for our purpose in this paper, since they are in general  solvable but highly non-local in the coordinate space.
In addition, the solvable potential is formally expanded in terms of derivatives, which can be compared with the derivative expansion in the HAL QCD method. We give basic properties of a separable potential we  consider in Sec.~\ref{sec:separable}.
We investigate how the derivative expansion works in the HAL QCD method for three cases in Sec.~\ref{sec:HALQCD}. The first one is to construct potentials from energy eigenfunctions. This is the cleanest case, where systematic errors for the derivative expansion are easy to estimate. In appendix~\ref{sec:appA}
coefficient functions in the derivative expansion of the potential are presented in this case. 
We compare them with those in the formal derivative expansion of the separable potential.
The second one is to evaluate potentials from time-dependent wave functions in the infinite volume, constructed as a solution of the time dependent Schr\"odinger equation with some initial condition.
While the finite volume method by definition does not work in this case, the time dependent HAL QCD method works to extract physical observables\cite{HALQCD:2012aa}. An issue in this case is how reliable results from the derivative expansion are.
We compare phase shifts obtained from the potentials at lowest few orders in the derivative expansion with the exact result. 
Finally, we consider a construction of the potential from time-dependent wave functions composed of a finite sum of eigenfunctions,
which mimic  time-dependent wave functions in the finite volume.
This is most similar to actual setups in lattice QCD simulations performed on a finite volume with a finite lattice spacing. 
We give our conclusion in Sec.~\ref{sec:conclusion}.
Details for the calculation of time-dependent wave functions are presented in appendix~\ref{sec:appB}
 
 The preliminary result on a similar analysis with a different separable potential can be found in \cite{Aoki:2019pnq,Aoki:2020bew}.

\section{Separable potential}
\label{sec:separable}
Let us consider the Schr\"odinger equation  with non-local potential, given by
\beqa
 (E_k - H_0) \psi_k(\vec x) &=& \int d^3y\,  V(\vec x, \vec y)  \psi_k(\vec y)=  \omega v(\vec x) \int d^3y\,  v^\dagger (\vec y) \psi_k(\vec y),
\eeqa
where
\beqa
E_k &:=&{\vec k^2\over 2m}, \quad H_0= -{\vec \nabla^2\over 2m},
\eeqa
and we take a separable potential $V(\vec x, \vec y) := \omega v(\vec x) v^\dagger (\vec y)$ in the last line.
For the general method to investigate scattering problems  with separable potentials, for example, see \cite{Augusiak:2005a}. 

The corresponding Lippmann-Schwinger equation becomes
\begin{eqnarray}
\psi_k(\vec x) &=& e^{i \vec k\cdot\vec x} -\omega \int d^3 y G_k(\vec x,\vec y) v(\vec y)
\int d^3z\, v^\dagger (\vec z) \psi_k(\vec z),  
\label{eq:LSE}
\end{eqnarray}
where the Green's function is given by
\begin{eqnarray}
G_k(\vec x,\vec y) &:=& \int \frac{d^3 p}{(2\pi)^3} \frac{2m e^{i \vec p (\vec x-\vec y)}}{\vec p^2-k^2-i\epsilon}
=\frac{m}{2\pi} \frac{e^{i k\vert \vec x-\vec y\vert}}{\vert \vec x-\vec y\vert} 
\end{eqnarray}
Eq.~\eqref{eq:LSE} can be solved as
\begin{eqnarray}
\psi_k(\vec x) &=&  e^{i \vec k\cdot\vec x} -  \langle \vec x \vert G_k \vert v\rangle
 \frac{\langle v \vert \vec k\rangle}{ \dfrac{1}{\omega} +  \langle v\vert G\vert v\rangle},\quad
 e^{i\vec k\vec x} = 4\pi \sum_{lm} i^l j_l(kr) Y_{lm}(\Omega_{\vec x})Y_{lm}(\Omega_{\vec k})^\dagger, 
\end{eqnarray}
where we define
\begin{eqnarray}
\langle v \vert \vec k\rangle &:=& \int d^3x v^\dagger(\vec x) e^{i \vec k\cdot\vec x} , \\
\langle \vec x \vert G_k \vert v\rangle &:=& \int d^3 y\, G_k(\vec x,\vec y) v(\vec y), \quad
 \langle v\vert G_k\vert v\rangle := \int d^3x \, v^\dagger(\vec x) \langle \vec x \vert G_k \vert v\rangle .
\end{eqnarray}

In this paper, we take one choice for $v(\vec x)$ as
\beqa
v(\vec x) &=& {e^{-\mu x}\over x} = v^\dagger(\vec x), \quad x=\vert\vec x\vert .
\label{eq:model}
\eeqa

\subsection{Explicit solutions}
Using the formula
\begin{eqnarray}
\frac{e^{i k \vert \vec x-\vec y\vert }}{\vert \vec x -\vec y\vert} &=& 4\pi i k\sum_{l=0}^\infty \sum_{m=-l}^l h^{(+)}_l (k r_>) j_l(k r_<) Y_{lm}(\Omega_{\vec x})
Y_{lm}^\dagger (\Omega_{\vec y}), 
\end{eqnarray}
where $r_>:= \max(x,y)$,  $r_<:=\min(x,y)$,  we obtain
\begin{eqnarray}
\langle v \vert \vec k\rangle &=& {4\pi\over \mu^2+k^2}, \
 \langle v\vert G\vert v\rangle = {8\pi m\over (\mu^2+k^2)^2}\left[{\mu^2-k^2\over 2\mu} + ik\right],\
\langle \vec x \vert G \vert v\rangle = {2m (e^{ikx} -e^{-\mu x})\over (\mu^2+k^2) x},~~~
\label{eq:vGv}
\end{eqnarray}
where we use 
\begin{eqnarray}
Y_{00}(\Omega_{\vec x}) :={1\over \sqrt{4\pi}}, \quad
h_0^{(+)}(z) &:=& -i \frac{e^{i z}}{z}, \quad j_0(z) := \frac{\sin z}{z} .
\end{eqnarray}

Thus scattering states exist only for the S-wave ($l=0$) as
\beqa
\psi_k^0(x) &=&{\sin (kx +\delta(k) ) -\sin\delta(k) e^{-\mu x}\over k x}
\label{eq:psi0}
\eeqa 
for $k\ge 0$, and the scattering phase shift becomes
\beqa
k \cot \delta(k) ={1\over a_0} + {r_{\rm eff}\over 2} k^2 + P_4 k^4,
\label{eq:ps0}
\eeqa
where the scattering length, the effective range and the shape parameter, respectively, are given by
\beqa
{1\over a_0} &=& -{\mu\over 2}\left[ 1+{\mu^3\over c}\right], \quad r_{\rm eff} = {1\over \mu}\left[1- {2\mu^3\over c}\right],\quad
P_4= -{1\over 2c}
\eeqa
with $c:=4\pi m\omega$.

For the bound state $\vert B\rangle$, eq.~\eqref{eq:LSE}  leads to
\beqa
\langle v\vert B\rangle &=& -\omega \langle v\vert G_k\vert  v\rangle  \langle v\vert B\rangle, \quad
\langle \vec x\vert B\rangle = -\omega \langle \vec x\vert G_k\vert  v\rangle  \langle v\vert B\rangle, 
\eeqa
which determines the binding momentum $k=i\gamma_B$ and the normalized bound state as
\beqa
\gamma_B = \sqrt{- c\over \mu} -\mu, \quad \langle \vec x\vert B\rangle = {\mu N_B \over 2\pi x (\mu-\gamma_B)}
\left(e^{-\gamma_B x} -e^{-\mu x}\right), \quad N_B^2:= {2\pi \gamma_B(\mu+\gamma_B)\over  \mu}.
\label{eq:bound}
\eeqa

\subsection{Infrared cut-off}
The effective range expansion (ERE) of the scattering phase shift in eq.\eqref{eq:ps0} is too simple, as it is  the 2nd order polynomial of $k^2$. In order to make the ERE of the phase shift a more complicated function of $k^2$, we introduce an infrared cut-off $R$ and
modify the wave function as
\beqa
\psi_k^R(x) &=& \left\{
\begin{array}{cc}
\psi_k^0(x)  & ( r < R)    \\
\\
  \ds C(k) {\sin ( k x +\delta_R(k) ) \over kx} & ( r\ge R)     \\
\end{array}
\right. ,
\label{eq:psiR}
\eeqa 
where the continuity of the wave function and its derivative at $x=R$ leads to
\beqa
C(k) &=& {X\over \sin (kR +\delta_R(k) )}, \quad k\cot \delta_R(k) = k{X +\cot(kR) Y\over \cot(kR) X -Y},
\label{eq:deltaR}
\\
X&=&\sin (kR +\delta(k)) -\sin\delta(k) e^{-\mu R}, \quad
Y= \cos (kR +\delta(k)) +{\mu\over k} \sin\delta(k) e^{-\mu R}.
\end{eqnarray}
Thus the scattering length $a_R$ is given by
\begin{equation}
a_R = a_0 {1-(1+\mu R) e^{-\mu R}\over 1+ a_0\mu e^{-\mu R}}.
\end{equation}
Note that an introduction of $R$ also modifies $\gamma_B$ and $\langle\vec x \vert B\rangle$, where
$\gamma_B$ in the presence of the infrared cutoff $R$ is estimated by an analytic continuation of $k \cot \delta_R(k)$, as will be shown later.

\subsection{Formal derivative expansion}
Using the Taylor expansion, we decompose the separable potential directly in terms of derivatives as
\beqa
V(\vec x,\vec y) &=& \sum_{n=0}^\infty V^{\mu_1\cdots\mu_n}_n(\vec x)  \partial^x_{\mu_1}\cdots\partial^x_{\mu_n} \delta^{(3)}(\vec x- \vec y)
\eeqa
where
\beqa
V^{\mu_1\cdots\mu_n}_n(\vec x) &=& {1\over n!}
\omega v(\vec x)   \int d^3y\, v(\vec y)(y-x)^{\mu_1}\cdots (y-x)^{\mu_n}  .
\eeqa
The lowest few orders corresponding to $V(\vec x)$ in \eqref{eq:model} are given by
\beqa
V_0(x) &=&  {4\pi\omega e^{-\mu x} \over \mu^2 x}, \
V_1^\mu (\vec x) =  -V_0(x) x^\mu, \
V_2^{\mu\nu}(\vec x) =   V_0(x) \left[ {\delta^{\mu\nu}\over \mu^2} + {x^\mu x^\nu\over 2}\right], 
\eeqa
Defining $V_n:= V_n^{\mu_1\cdot \mu_n}\partial_{\mu_1}\cdots \partial_{\mu_n}$, we obtain 
\beqa
V_1\left(x,{d\over dx}\right) &=& V_0(x)-V_0(x){d\over d x}x, \nn
V_2\left(x,{d\over dx}\right) &=& V_1\left(x,{d\over dx}\right) +V_0(x)\left({1\over \mu^2} +{x^2\over 2}\right){1\over x}{d^2\over dx^2} x ,
\eeqa

By introducing the infra-red cutoff $R$ again as $V_0(x) \to V_0(x)\theta(R-x)$ in the above expressions, we calculate the scattering phase shifts with the potential $U_n:=\ds\sum_{i=0}^n V_i$  for $n=0,1,2$. 
In the presence of the infra-red cutoff, the exact phase shift, denoted by $\tilde \delta_R(k)$, is given by
\beqa
k \cot \tilde \delta_R(k) &=& -k {{\rm Re} (1- S(k) )\over {\rm Im}(1-S(k))},  
\label{eq:deltaRt}
\eeqa
where
\beqa
S(k) := {c \over \mu( k+i\mu)^2}\left( 1- {e^{-\mu R}\{ e^{-\mu R}(k+i\mu) -2i \mu e^{i kR}\}\over k-i\mu}\right),
\eeqa
which is obtained by replacing $\vert v\rangle\to \vert v_c\rangle$  with $v_c(\vec x) := \theta(R-x) v(\vec x)$
in \eqref{eq:vGv}. The resulting potential is no longer Hermitian but the on-shell T-matrix satisfies the unitarity and the scattering can be described by a real phase shift, $\tilde{\delta}_{R} (k)$. Although $\tilde \delta_R(k)$ and $\delta_R(k)$ in \eqref{eq:deltaR} are different in their expressions,
they are almost identical in numbers as long as $R$ is reasonably large.

\begin{figure}[tbh]
 \centering
\includegraphics[width=0.49\textwidth,clip]{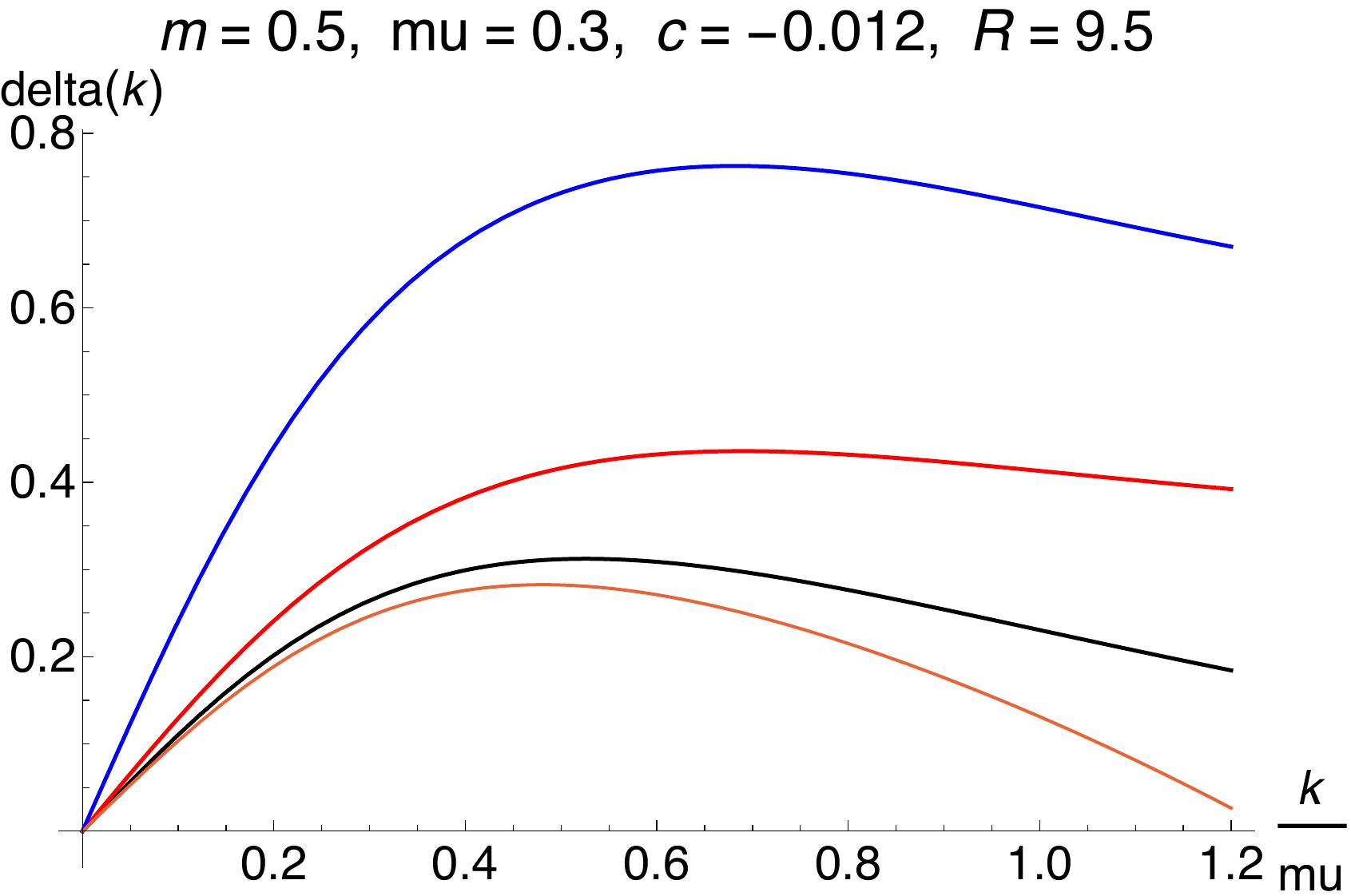}
 \includegraphics[width=0.49\textwidth,clip]{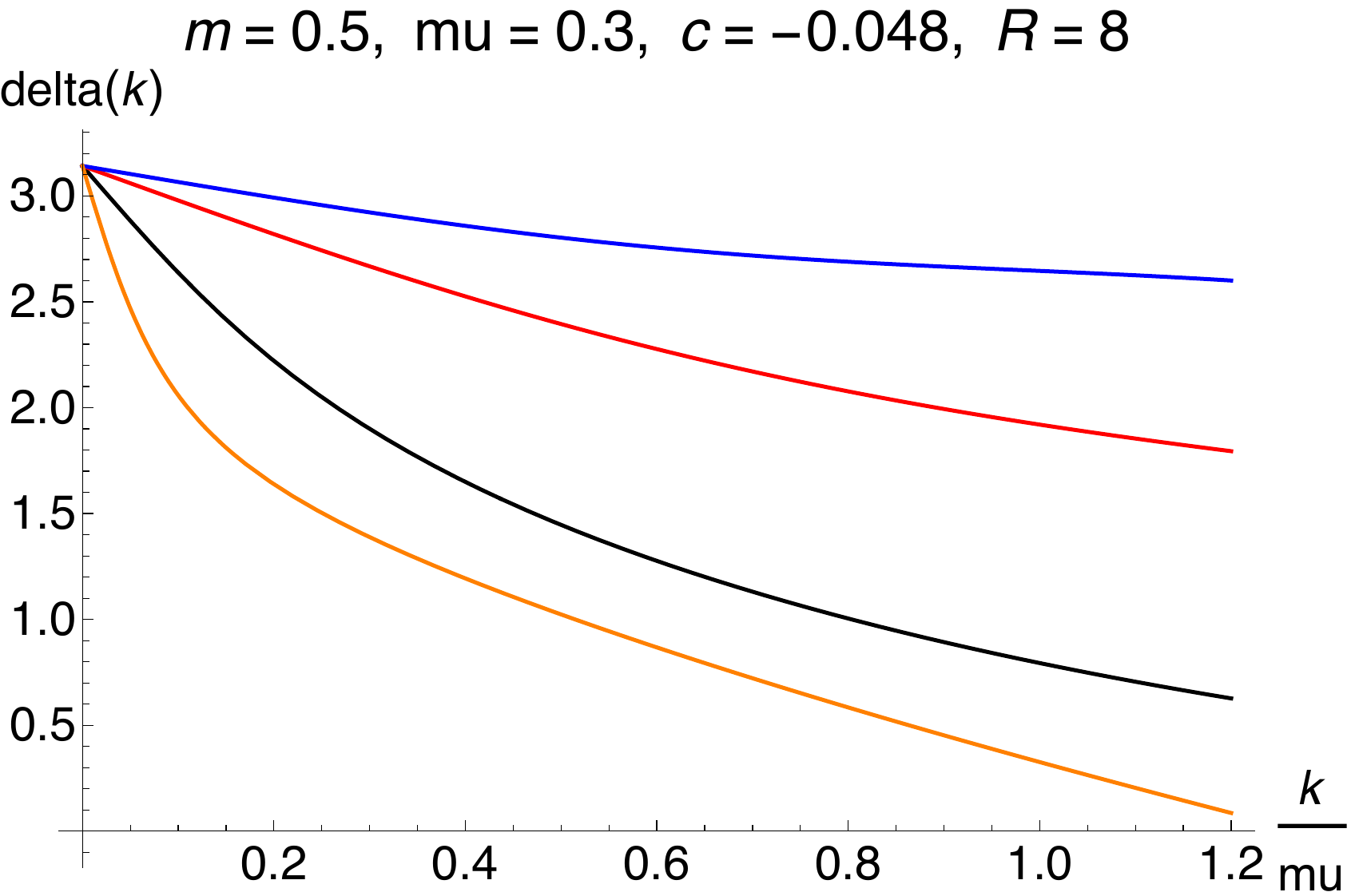}
  \caption{
    \label{fig:PhaseShift}
     Scattering phase shift calculated with the potential in the formal derivative expansion are plotted as a function of $\dfrac{k}{\mu}$
     with $m=0.5$ and $\mu=0.3$.   Results with $U_0 $, $U_1$ and $U_2$ are shown by red, blue and orange lines, respectively, together with the exact one $\tilde\delta_R(k)$ by black line. 
We take $c=-0.012$ and $R=9.5$, which allows no bound state (Left), while $c=-0.048$ and $R=8$, which produces one bound state (Right).       
        }
\end{figure}

Fig.~\ref{fig:PhaseShift} shows the scattering phase shift with $m=0.5$ and $\mu=0.3$.
In the left figure, we take $c=-0.012$ and $R=9.5$, which is attractive but without bound states, while in the right, we take $c=-0.0048$ and $R=8$  to have one bound state, whose binding energy is given by $\gamma_B^2\simeq 0.00570$.
In the figure,  red, blue and orange lines represent the phase shift obtained with $U_0$, $U_1$ and $U_2$, respectively,
together with the exact one $\tilde\delta_R(k)$ in \eqref{eq:deltaRt} by the black line for a comparison.
In both cases (with and without bound state),
while $U_0$ reasonably approximate the behavior at low energies, $U_1$ worsens  but $U_2$ improves the agreement\footnote{We also confirm that $U_3$ and $U_4$ do not improve approximations at all.} .
The approximation is a little better for $c=-0.012$.
The binding energy is approximated as $\gamma_B^2\simeq 0.0685$ ($U_0$), 0.384 ($U_1$), and 0.000471 ($U_2$).

In the next section, we consider  the HAL QCD method,
and we reconstruct the potential in a derivative expansion of a similar kind, using behaviors of wave functions
under the separable potential  \eqref{eq:model} .

\section{Derivative expansion in the HAL QCD method}
\label{sec:HALQCD}

\subsection{Potential from eigenfunctions}
In this subsection, we assume that $n$ eigenfunctions $\psi^R_k$ in \eqref{eq:psiR} are available for us to construct the potential, which thus satisfies 
\beqa
(E_{k_i} - H_0) \psi_{k_i}^R(x) = V_n\left(x,{d\over dx} \right) \psi_{k_i}^R(x),\quad i=0,1,\cdots, n ,
\label{eq:Sch_hal}
\eeqa
where $V(\vec x, \nabla)$ are replaced by $V_n(x, {d\over d x})$, since
the scattering occurs only in S-wave by the separable potential defined with \eqref{eq:model} .  

Eq.~\eqref{eq:Sch_hal} for $n+1$ eigenfunctions determines $n+1$ independent local functions in $V_n(x,\nabla)$
for $n=0,1,2,\cdots$,
which is taken as
\beqa
V_n(x,\nabla) &=& \sum_{i=0}^{n} V_{n,i}(x) (\nabla^2)^i, \quad \nabla^2 ={1\over x}{d^2\over d x^2} x,
\label{eq:hal_exp}
\eeqa
where the absence of odd derivative terms is our choice for a {\it scheme} to define potentials in the HAL QCD method.
Although it is certainly possible to take another scheme including odd derivative terms for the potential,
we think that our scheme without them is more efficient.
Since odd derivative terms are absent in the hermitian potential 
with rotational and time-reversal symmetries,
we do not need such terms to describe scattering phase shift.
Indeed the first derivative term in the formal derivative expansion worsen the approximation, as seen in the previous section. 

Eq.~\eqref{eq:Sch_hal} leads to $V_{n,i}$ as
\beqa
\sum_{j=0}^{n}T_{ij}(x)  V_{n,j}(x) &=& K_i(x) \Rightarrow  V_{n,i}(x) = \sum_{j=0}^{n}\left[T^{-1}(x)\right]_{ij} K_j(x), 
\eeqa
where
\beqa
T_{ij}(x)  := {1\over x} {d^{2 j}\over d x^{2j}} \{x\psi^R_{k_i}(x)\} , \quad
K_i(x) := {1\over 2m}\left( k_i^2 + {1\over x} {d^2\over dx^2}x\right)\psi_{k_i}^R(x).
\eeqa
Note that $V_n$ is an approximated potential, which depends on the choice of $k_i$ ($i=0,1,2,  \cdots, n$) as
it gives correct results only at $k_i$ ($i=0,1,2,  \cdots, n$).
In this subsection, we show results with $n=0$ (LO), $n=1$ (NLO) and $n=2$ (NNLO).

\begin{figure}[tbh]
  \centering
\includegraphics[width=0.49\textwidth,clip]{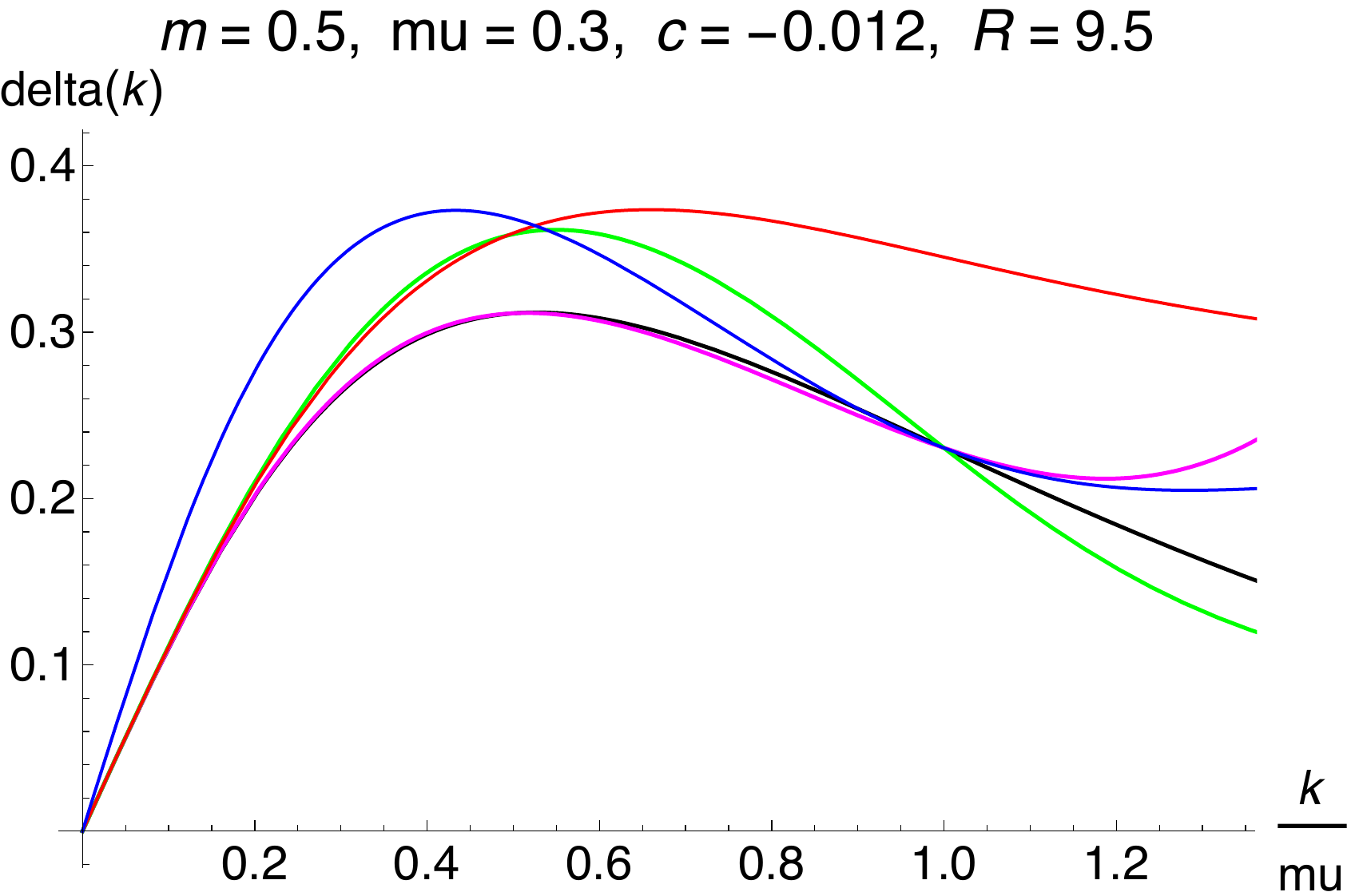}
 \includegraphics[width=0.49\textwidth,clip]{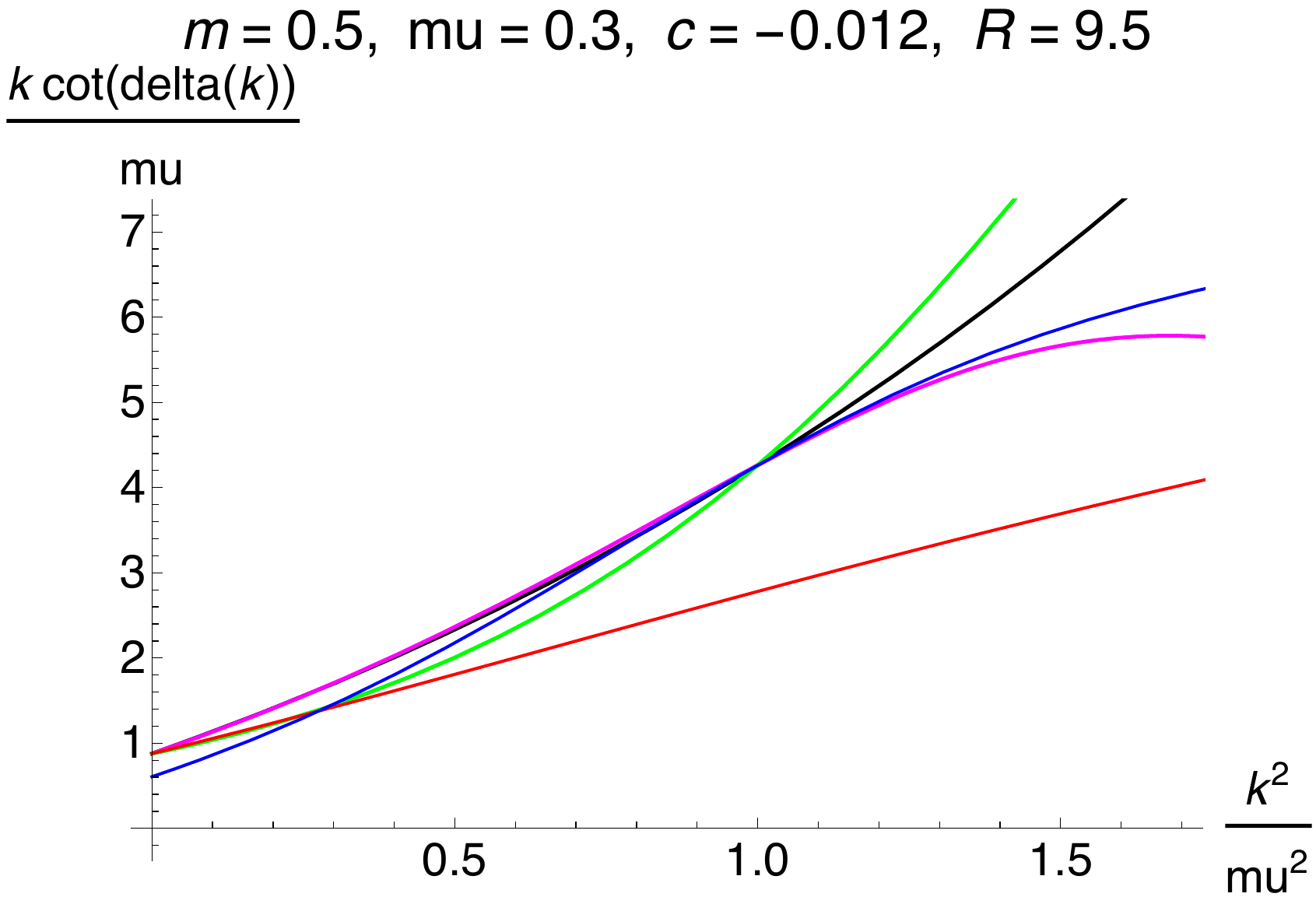}
  \caption{
    \label{fig:Eig_c12}
    (Left) Scattering phase shift  $\delta(k)$ as a function of $\dfrac{k}{\mu}$ at $m=0.5$, 
    $\mu=0.3$, $c~-0.012$ and $R=9.5$.
     The LO results from eigenfunctions at $k=0$ and $k=\mu$ are plotted by red and blue lines, respectively,
     while NLO and NNLO results are given by green and magenta lines, respectively, together with the exact results  $\delta_R(k)$ by the black line.  
      (Right) The corresponding $\dfrac{k}{\mu}\cot \delta(k)$ as a function of $\dfrac{k^2}{\mu^2}$ at same parameters. 
        }
\end{figure}

Fig.~\ref{fig:Eig_c12} represents the scattering phase shift $\delta (k)$ as a function of $\dfrac{k}{\mu}$ (Left) and $\dfrac{k}{\mu}\cot\delta(k)$ as a function of $\dfrac{k^2}{\mu^2}$ (Right) at $m=0.5$, $\mu=0.3$, $c=-0.012$ and $R=9.5$, which produces no bound state.
The LO result from the eigenfunction at $k=0$ (low energy) by the red line correctly reproduce the exact one by the black line at $k=0$. The LO result from the eigenfunction at $k=\mu$ (high energy) by the blue line, on the other hand, agrees with
the exact one at $k=\mu$.  
Pretending $\mu$ in $V(\vec x)$ as a mass of exchange particle, we may regard $k\simeq \mu$ as the threshold of the inelastic scattering in quantum field theory.
We obtain the NLO result, plotted by green line, by using two eigenfunctions, which by definition agree with exact ones at $k=0$ and $k=\mu$, and give a reasonable approximation at energy range between the two. 
Adding the third eigenfunction at $k=\mu/2$,  we can calculate the NNLO result, which is nearly exact from $k=0$ to $k=\mu$, as shown by the magenta line in the figure.  
This analysis demonstrates how the derivative expansion works in the HAL QCD method.
In contrast to the formal expansion, the HAL QCD method can incorporate information from the eigenfunction at high energy 
to improve the accuracy of the approximation. 

In appendix \ref{sec:appA}, we plot coefficient functions of the potential for these parameters as a function of $x$.

\begin{figure}[tbh]
  \begin{center}
\includegraphics[width=0.49\textwidth,clip]{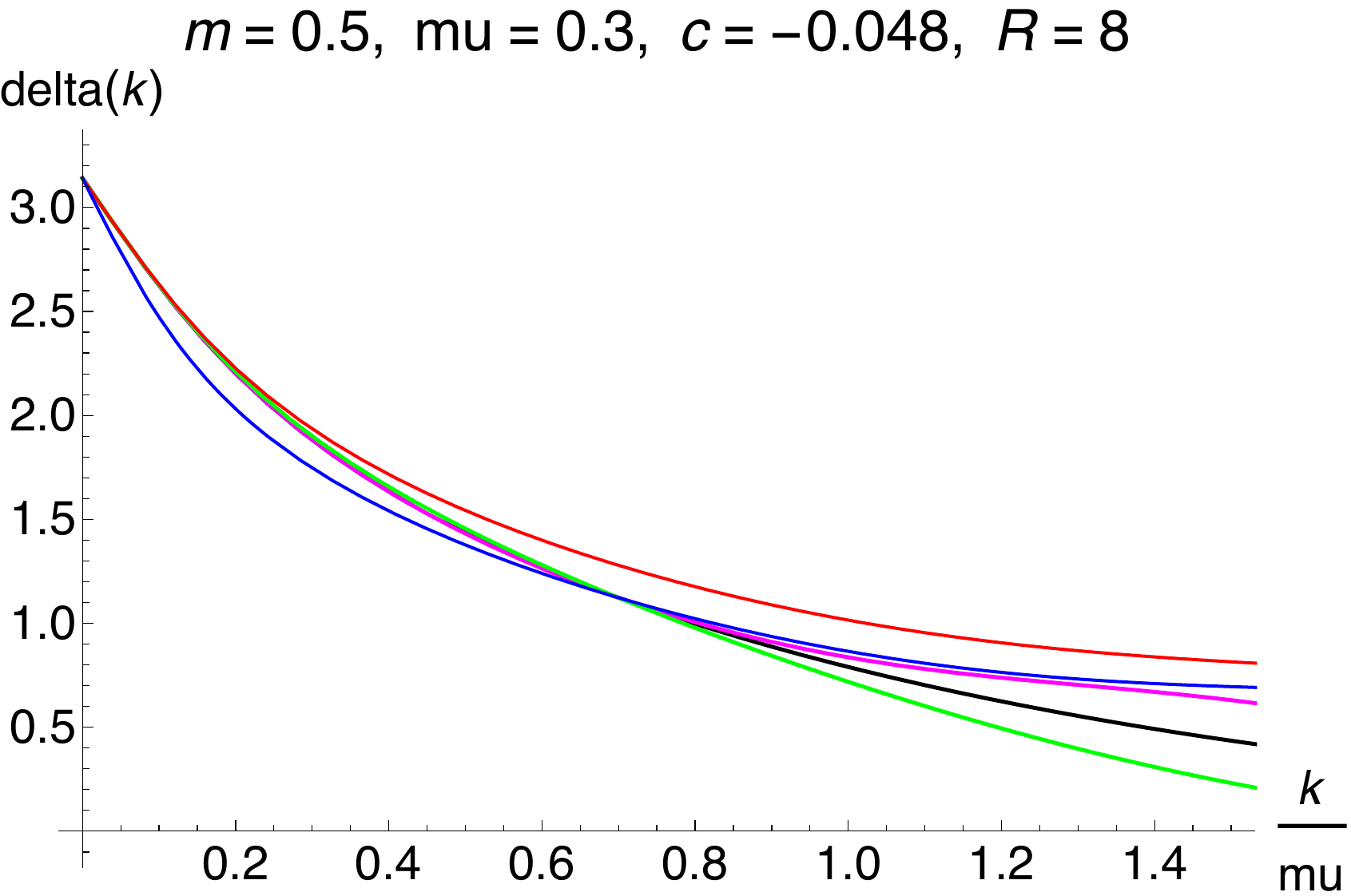}
 \includegraphics[width=0.49\textwidth,clip]{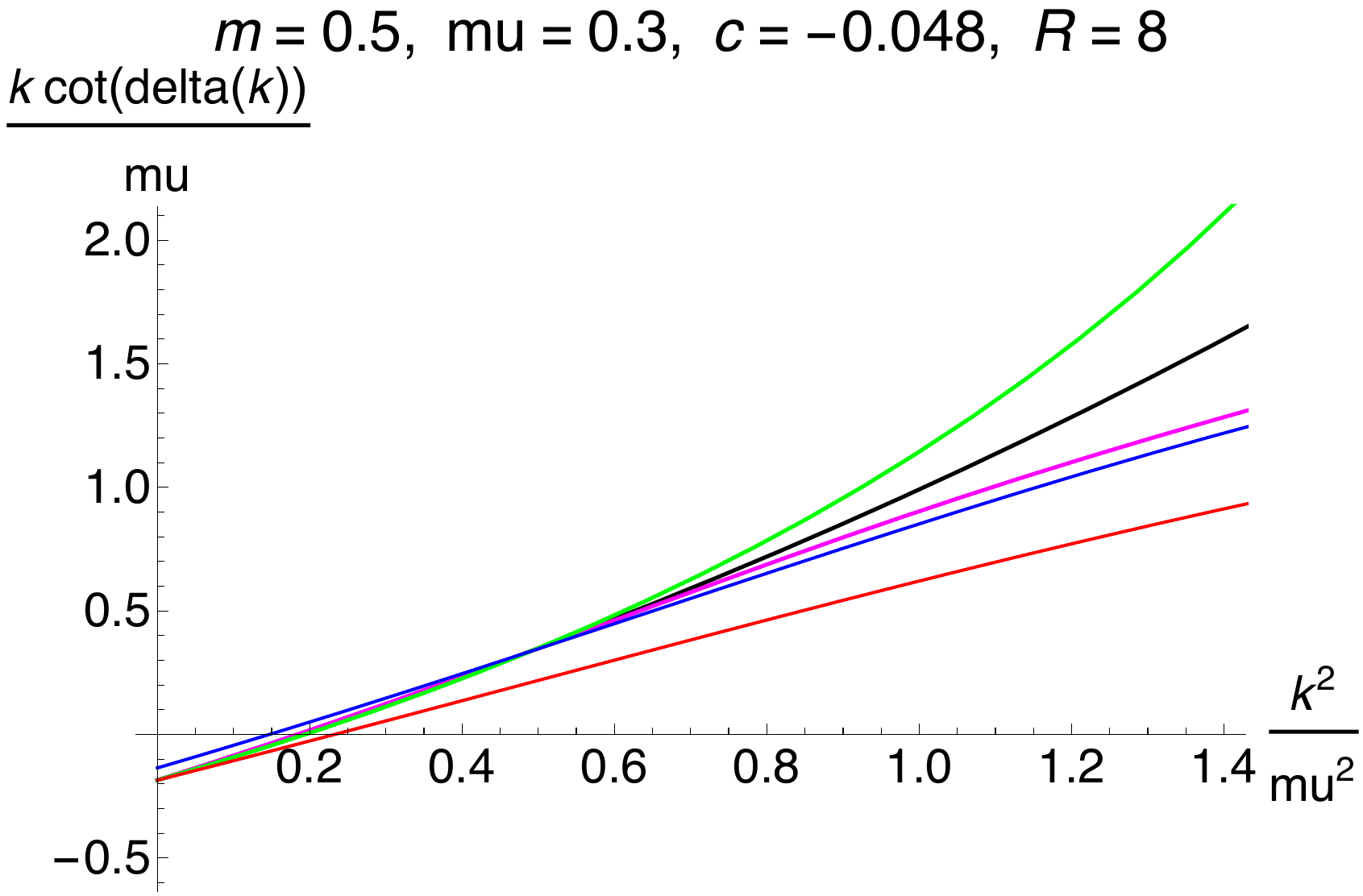}
 \vskip 0.5in
  \includegraphics[width=0.6\textwidth,clip]{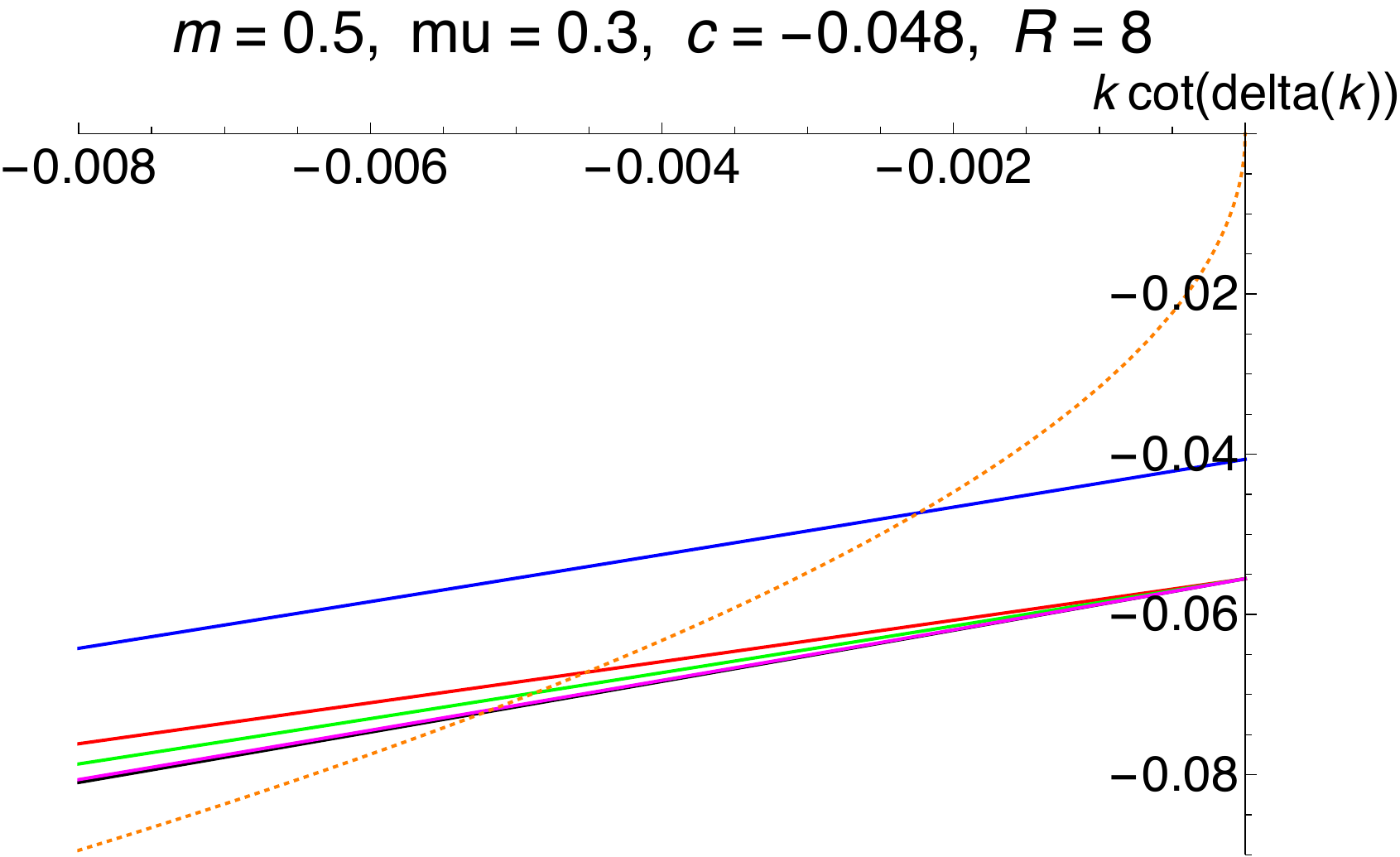}
   \end{center}
  \caption{
    \label{fig:Eig_c48}
    (Upper-Left) Scattering phase shift  $\delta(k)$ as a function of $\dfrac{k}{\mu}$ at $m=0.5$, 
    $\mu=0.3$, $c=-0.048$ and $R=8.0$.
     The LO results from eigenfunctions at $k=0$ and $k=0.7\mu$ are plotted by red and blue lines, respectively,
     while NLO and NNLO results are given by green and magenta lines, respectively, together with the exact results  $\delta_R(k)$  by the black line.  
      (Upper-Right) The corresponding $\dfrac{k}{\mu}\cot \delta(k)$ as a function of $\dfrac{k^2}{\mu^2}$ at same parameters. 
      (Lower) $k\cot \delta(k)$  for $k^2<0$, together with the bound state condition $-\sqrt{-k^2}$ by the orange dotted line.
        }
\end{figure}

Fig.~\ref{fig:Eig_c48} shows the scattering phase shift $\delta (k)$ as a function of $\dfrac{k}{\mu}$(Upper-Left) and $\dfrac{k}{\mu}\cot\delta(k)$ as a function of $\dfrac{k^2}{\mu^2}$ (Upper-Right) at $m=0.5$, $\mu=0.3$, $c=-0.048$ and $R=8.0$, which produces one bound state. 
As before, red and blue lines  represent  the LO results from the eigenfunctions at $k=0$ (low energy) and $k=0.7\mu$ (high energy),
respectively, 
while the green line gives the NLO result from two eigenfunctions at $k=0,0.7\mu$.
Finally, we obtain the NNLO result (magenta line), by adding the third eigenfunction at $k=0.35\mu$.  
As seen from the figure,
the NLO (green) and the NNLO (magenta) results agree with the exact one (black line) between $k=0$ and $k=0.7\mu$ 
at these parameter.
The derivative expansion in the HAL QCD method works well also for the coupling strong enough to have a bound state.

The lower figure shows analytic continuations of $k \cot \delta(k)$ to $k^2<0$, 
where  meanings of colors are same as in other figures while
the orange dotted line represents the bound state condition, $ -\sqrt{-k^2}$  as a function of $k^2<0$.
Note that $k$ is NOT normalized by $\mu$ in this figure.
An existence of an intersection $k_0^2$ between $k \cot \delta(k)$ and $-\sqrt{-k^2}$
means an existence of a bound state  whose binding energy is given by ${-k_0^2\over 2m}$.
As we increase the order of the expansion, LO(red), NLO(green), and NNLO (magenta),
the intersection moves toward the exact one, $k_0^2 = -\gamma_B^2 = -0.0052$, 
estimated by an intersection between $k\cot \delta_R(k)$ (black solid line) and $-\sqrt{-k^2}$ (orange dotted line).
which is nearly reproduced by the the NNLO result. 
Note that we do not include the wave function for the bound state to construct potentials.
The eigenfunction at low energy at $k\simeq 0$, in some sense, knows information of the bound state.
Interestingly, the LO result (blue line) from the eigenfunction at high energy ($k=0.7\mu$) leads to the worst result among all, 
$\gamma_B^2\simeq 0.0022$, a factor 2.6 smaller than the exact value, $\gamma_B^2\simeq 0.0057$.

\subsection{Potential from correlation functions in the time dependent HAL QCD method}
Since  it is not so easy to obtain each eigenfunction separately from correlation functions, which are linear combinations of eigenfunctions, the time dependent HAL QCD method\cite{HALQCD:2012aa} has been proposed to extract the potential directly from correlation functions
without decomposing them into eigenfunctions.
In this subsection, we apply the derivative expansion in the time dependent HAL QCD method to extract the potential from correlation functions.

The time dependent correlation function is defined by
\beqa
R(t,\vec x) &=& \int {d^3k\over (2\pi)^3} e^{-E_k t} f(\vec k) \psi_k (\vec x) + f_B e^{-E_B t} \langle \vec x\vert B\rangle,
\label{eq:Rcorr}
\eeqa  
where 
\beqa
E_k^\lambda &=& {k^2\over 2m}, \quad E_B^\lambda =- {\gamma_B^2\over 2m} ,
\eeqa
and $f(\vec k)$ and $f_B$ are determined so as to satisfy a given initial condition as $R(0,\vec x) = {\sigma^2 e^{-\sigma x} \over 4\pi x} $ with a parameter $\sigma$, which leads to $R(0,\vec x) =\delta(\vec x)$ in the $\sigma\to\infty$ limit. 
Details of calculations for $R$ and its derivative are presented in appendix~\ref{sec:appB}.

Suppose that we prepare $n+1$ independent correlation functions by taking $n+1$ different $\sigma$, denoted as $R_{\sigma_i}(t,x)$ ($i=0,1,\cdots, n$)
since they are functions of $x=\vert\vec x\vert$. As before $n$ local terms are extracted as
\beqa
V_{n,i}(t,x) &=& \sum_{j=0}^{n} \left[T^{-1}(t,x)\right]_{ij} K_j(t,x),
\eeqa
where
\beqa
T_{ij}(t,x) &:=& {1\over x}{d^{2j}\over x^{2j}} \left\{x R_{\sigma_j}(t,x)\right\}, \quad
K_j(t,x) = \left( -{\partial \over \partial t} + {1\over x}{d^2 \over dx^2} x\right) R_{\sigma_i}(t,x).
\eeqa
We then introduce the infrared cutoff as $V_{n,i}^R(t,x) = \theta(R-x)  V_{n,i}(t,x) $.
As in  the previous subsection, we calculate the scattering phase shifts for $n = 0$ (LO), 1 (NLO) and 2 (NNLO),
and compare them with the exact one.

\begin{figure}[tbh]
  \centering
\includegraphics[width=0.49\textwidth,clip]{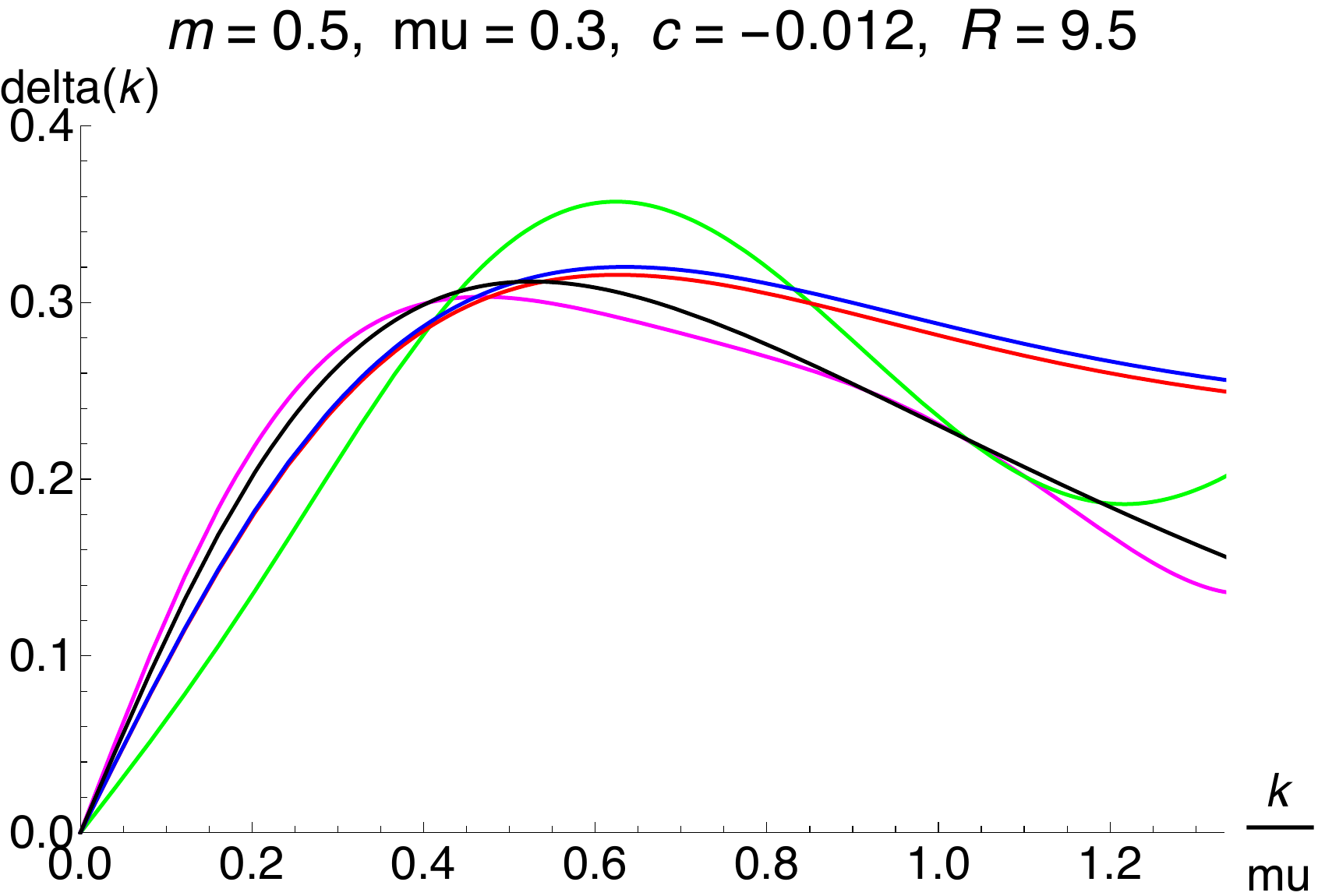}
 \includegraphics[width=0.49\textwidth,clip]{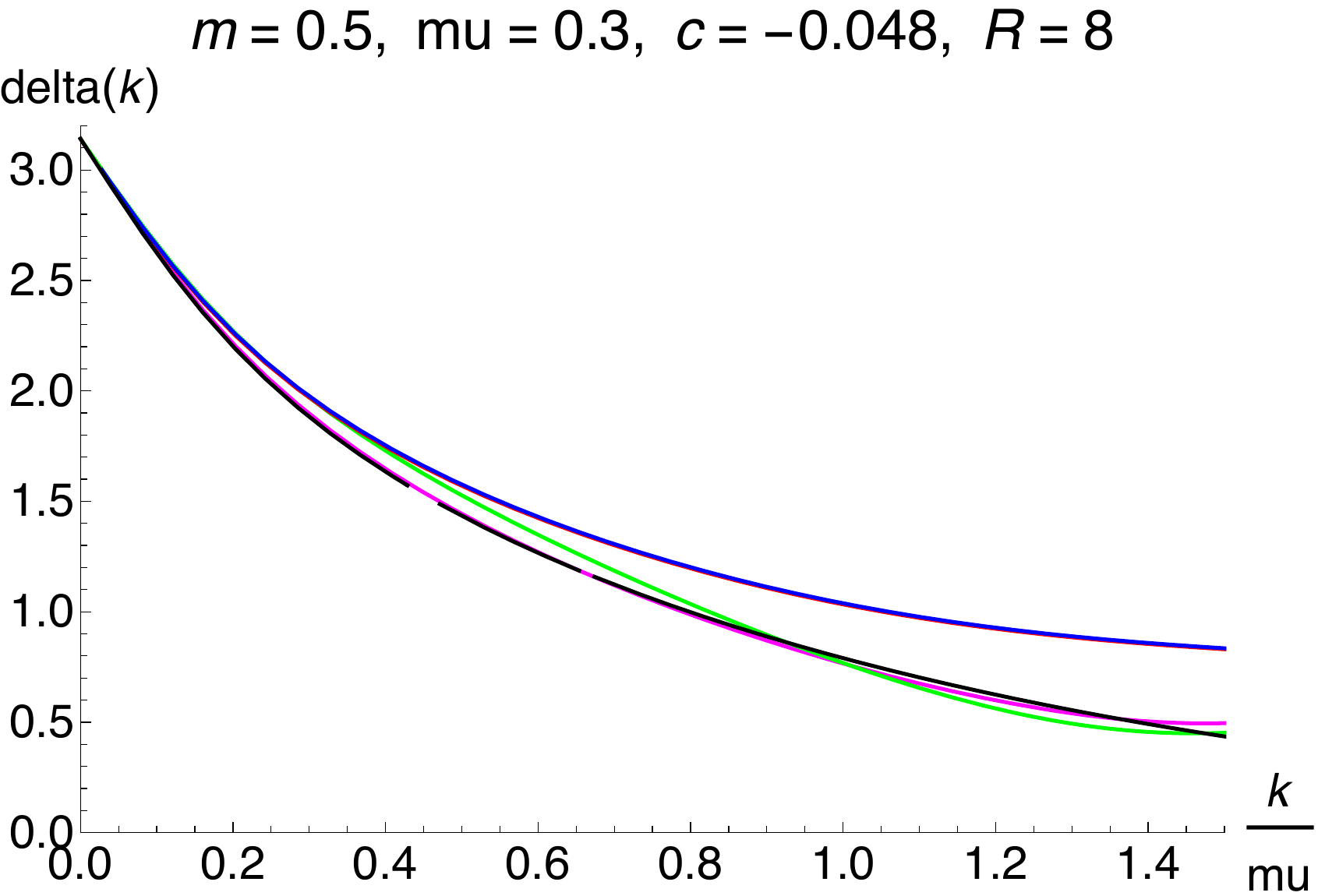}
  \caption{
    \label{fig:Rsig}
    Scattering phase shift  $\delta(k)$ as a function of $\dfrac{k}{\mu}$ at $m=0.5$ and $\mu=0.3$.
    (Left) $c=-0.012$ and $R=9.5$. (Right)  $c=-0.048$ and $R=8$. 
     The LO results from $R_\sigma(t=28,x)$ at $\sigma=\infty$ and $\sigma=0.3$ are plotted by red and blue lines, respectively,
     while NLO and NNLO results are given by green and magenta lines, respectively, together with the exact results  $\delta_R(k)$ by the black line.  
        }
\end{figure}
Fig.~\ref{fig:Rsig} show the scattering phase shifts $\delta(k)$ as a function of $\dfrac{k}{\mu}$
at $m=0.5$ and $\mu=0.3$. 
We take $c=-0.012$ ($c=-0.048$) and $R=9.5$ ($R=8$) in the left (right), where
the LO results from $R_\sigma(t,x)$ at $\sigma=\infty$ and $\sigma=0.3$ with $t=28$ are denoted by red and blue lines, respectively, 
while the NLO from the both by  the Green line. One more additional $R_\sigma(t,x)$ at $\sigma=0.6$ leads to the NNLO result by the magenta line, and the black line represents the exact one, $\delta_R(k)$.

As seen from the figures, the LO results reproduce the exact one at low energy at $k\simeq 0$.
At $k \le 1.2\mu$, which is a little larger than $\mu$, the LO results at $\sigma=\infty$ (red) and $\sigma=0.3$ (blue) are not so different,
and are almost identical for $c=-0.048$ (Right).
Combining these two, we obtain the NLO (Green), which is not so much better than the LOs for $c=-0.012$ (Left), but is certainly better than the LOs for $c=-0.048$ (Right). In both cases, the NNLO result nearly reproduces the exact  one
between $k=0$ and $k=\mu$. In particular, the agreement is excellent for $c=-0.048$. 
This indicates that the derivative expansion in the (time-dependent) HAL QCD method can be applied not only to eigenfunctions but also to $t$ dependent correlation functions.

\subsection{Potential from correlation functions with finite volume spectra}

Since lattice QCD simulations are usually performed in a finite box with the finite lattice spacing, energy eigenvalues are discrete
and bounded from above.
Thus the integral over $\vec k$ in \eqref{eq:Rcorr} becomes a summation over discrete momentum with the ultra-violate cutoff.
It is natural to ask how this discrete summation for the definition of $R$ affects the previous analysis for the derivative expansion in the HAL QCD method.
Since it is difficult to solve the Sch\"odinger equation in a finite box analytically, however, 
we emulate a similar situation replacing the integral in   \eqref{eq:Rcorr} with a finite discrete summation by hand.

Explicitly, we construct an $S$-wave correlation function as a sum over discrete momenta, $\vec{k}_{\vec{\nu}} = 
\frac{2\pi}{L} \vec{\nu} = \frac{2\pi}{L} (\nu_{1}, \nu_{2}, \nu_{3})$ with $\nu_{i} = 0, \pm1, \pm 2, \cdots$. Defining 
$n \equiv \vec{\nu}^{2} = \nu_{1}^{2} + \nu_{2}^{2} + \nu_{3}^{2}$, we can express the correlation function as 
\beqa
R_{L}(x; \tau, s) = 
\sum_{n=0} ^{N}  c(\tau,s,n) \psi_{k_{n}}^0(x),
\eeqa
where 
\beqa
c(\tau,s, n) &=& w(n) s^{n} e^{ -  \tau E_n}, \quad  E_n :={k^2_n\over 2m},
\quad
k_n := {2\pi\over L}\sqrt{n},
\eeqa 
$L$ is a spatial extension of the box, 
$w(n)$ is a number of the states whose energy is $E_n$.
 Parameters $\tau$ and $s=\pm 1$ control the size and the sign of  the coefficient for each state.
In this subsection, we consider three correlation functions with
(a) $(\tau,s)=(5, 1)$,  (b) $(\tau,s)=(20, -1)$ and  (c) $(\tau,s)=(40, 1)$, taking $L=48$ and $N=4$ for all cases.
Table~\ref{tab:param} shows $w(n)$ and $c(\tau,s,n)$ for each case.
The size of maximum momentum gives  $k\simeq 0.26$.
Note that the bound state is not included.

\begin{table}
\begin{center}
\begin{tabular}{|c| cc | c | c| c |}
\hline
$n$ & $w(n)$ & $k_n$ & \multicolumn{3}{c|}{ $c(\tau,s, n)$} \\
\hline
  & & &(a) $(\tau,s)=(5,1)$& (b) $(\tau,s)=(20,-1)$ & (c) $(\tau,s)=(40,1)$\\
\hline
0 & 1 & 0.0& 1.0 & 1.0 & 1.0\\
1 & 6 & 0.1309&5.50736& -4.25913  & 3.02336 \\
2 & 12 & 0.18512 & 10.1103 &6.04673 & 3.04691\\
3 & 8 & 0.226725 & 6.18682 & -2.86153 & 1.02355 \\
4 & 6  &0.261799 & 4.25913 & 1.52346 & 0.386819 \\
\hline
\end{tabular}
\end{center}
\caption{Parameters for three wave functions.}
\label{tab:param}
\end{table}
 
 \begin{figure}[tbh]
  \begin{center}
\includegraphics[width=0.49\textwidth,clip]{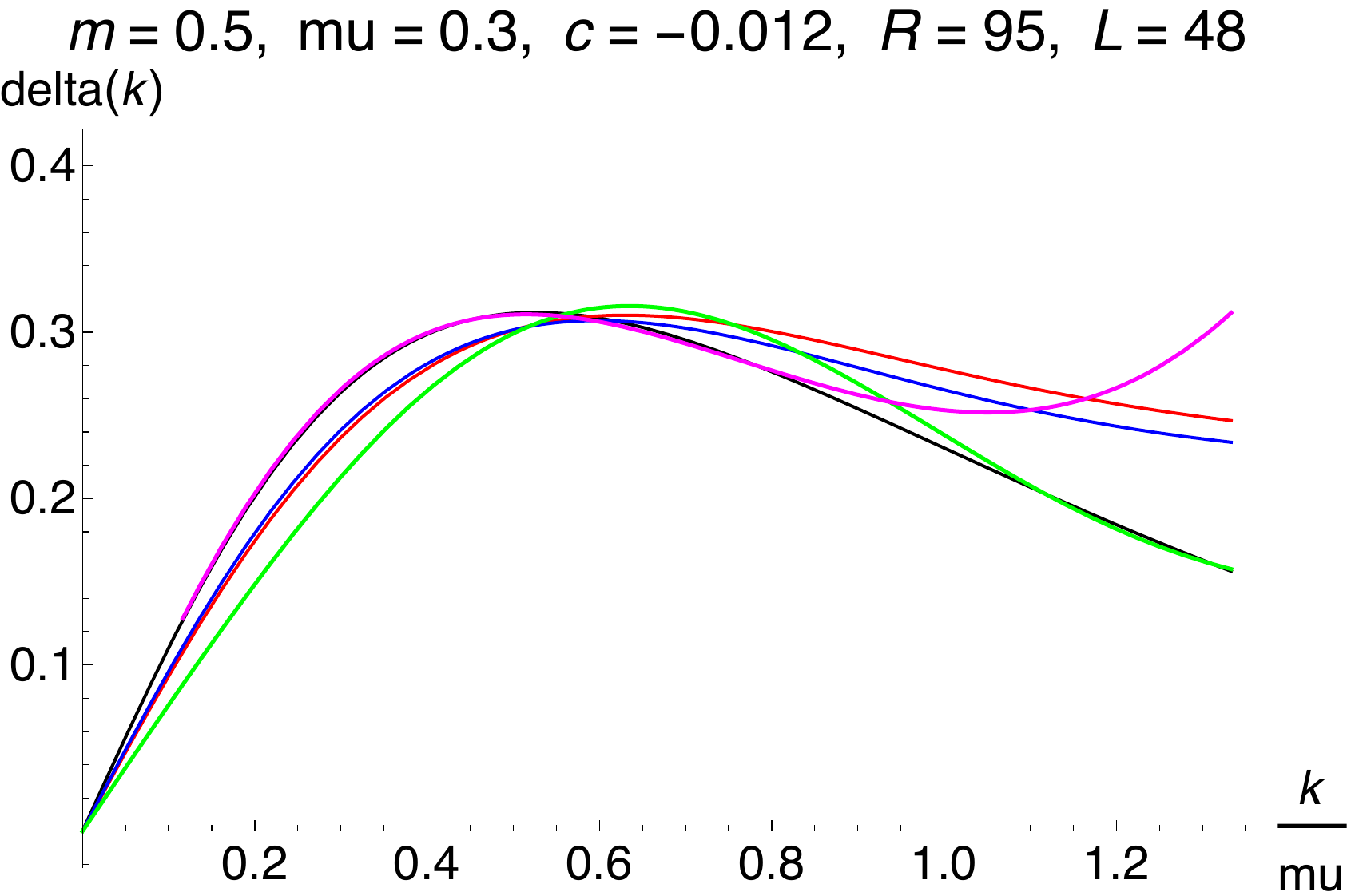}
 \includegraphics[width=0.49\textwidth,clip]{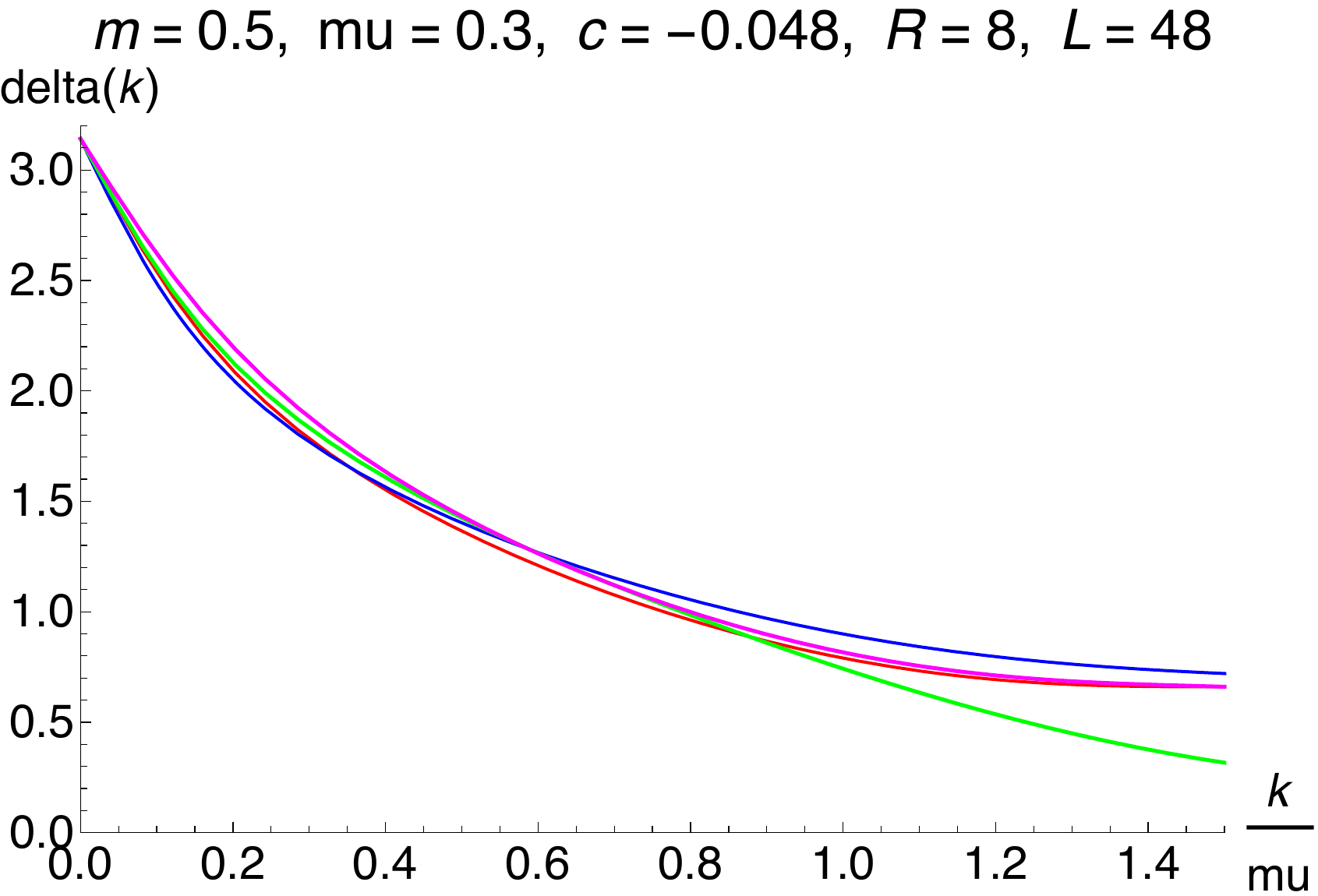}
 \end{center}
  \caption{
    \label{fig:PhaseShiftFV}
    Scattering phase shift  $\delta(k)$ as a function of $\dfrac{k}{\mu}$ at $m=0.5$, $\mu=0.3$ with $L=48$ and $N=4$.
    (Left) $c=-0.012$ and $R=9.5$.  (Right)  $c=-0.048$ and $R=8$. 
    In both figures,
    the LO results are plotted by red ((a) $(\tau,s)=(5,1)$) and blue ((b) $(\tau,s)=(20,-1)$) lines,
    while NLO and NNLO results are given by green and magenta lines, respectively, together with the exact results  $\delta_R(k)$ 
    by the black line. To obtain the NNLO result, we combine one more $R_L$ ( (c) $(\tau,s)=(40,1)$) with the previous two. 
        }
\end{figure}

Fig.~\ref{fig:PhaseShiftFV} represents scattering phase shift $\delta(k)$ as a function of $\dfrac{k}{\mu}$ at $m=0.5$ and $m=0.3$ with $L=48$
and $N=4$.
In the left (right) figure, we take $c=-0.012$ and $R=9.5$ ($c=-0.048$ and $R=8$) as before, where the LO results are obtained from $R_L(x;\tau,s)$
at (a) $(\tau,s)=(5,1)$ (red) and (b) $(\tau,s)=(20,-1)$ (blue), while the NLO from the both 
and the NNLO from these two plus an additional   $R_L(x;\tau,s)$ at (c) $(\tau,s)=(40,1)$ are shown by green and magenta lines, respectively.

As seen from left figure, the LO results from $R_L(x;\tau,s)$ both at (a) (red) and (b)  (blue) roughly reproduces the overall behavior of the exact phase shift (black) but the agreements are not so good. 
While the NLO (green) improve the agreement  a little,
the NNLO (magenta) almost reproduces the exact phase shift at $k \le 0.8\mu$, 
but deviates a lot from the exact one at $k > \mu$.
This is understandable, since the correlation functions $R_L$ do not contain any states with $k > 0.9\mu$. 
From this point of view, the agreement between the NLO (green)  and the exact one (black)  at $k > 0.9\mu$
is unexpected. It might be accidental or states with $k<0.9\mu$ might ``know"  $\delta_R(k)$ at $k > 0.9\mu$. 
A similar but much milder tendency is also found in the right figure.

\begin{figure}[tbh]
  \begin{center}
 \includegraphics[width=0.6\textwidth,clip]{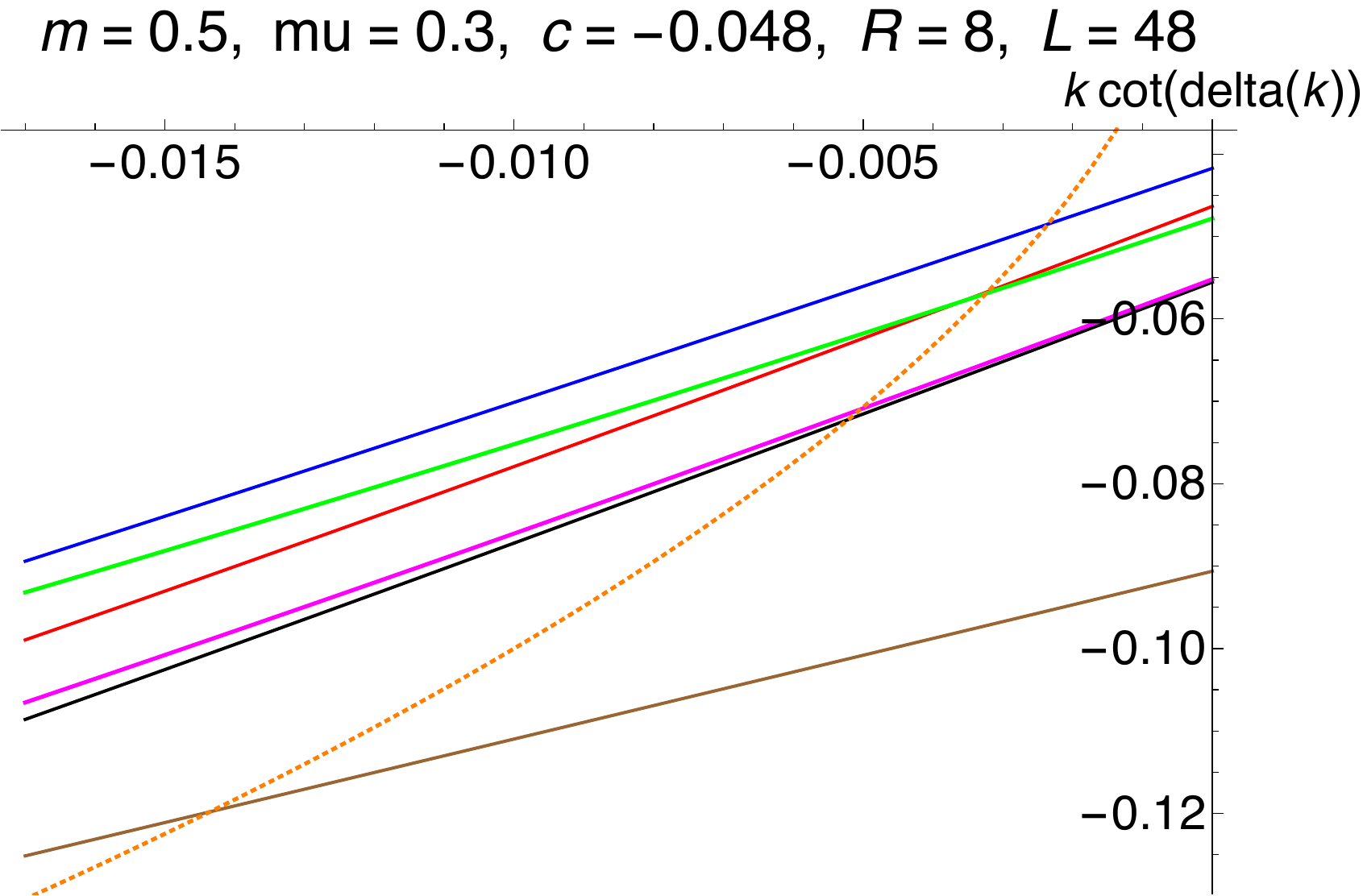}
 \end{center}
  \caption{
    \label{fig:kcotdFV}
    $k\cot \delta(k)$ as a function of $k^2 < 0$ at $m=0.5$, $\mu=0.3$, $c=-0.048$, $R=8$, $N=4$ and $L=48$.
    The LO results are plotted by red (a), blue(b) and brown (c) solid lines,
    while NLO and NNLO results are given by green and magenta solid lines, respectively, together with the exact results  
    $k\cot \delta_R(k)$ by the black solid line and the bound state condition $-\sqrt{-k^2}$ by the orange dotted line. 
            }
\end{figure}
In Fig.~\ref{fig:kcotdFV}, we plot $k\cot \delta(k)$ as a function of $k^2$ for $k^2<0$ at $c=-0.048$ keeping $m,\mu, R, L$
same as before,
together with the bound state condition $-\sqrt{-k^2}$, so that the crossing point between $k\cot\delta(k)$ and  $-\sqrt{-k^2}$ 
gives the binding energy.
Let us first consider the LO results, plotted by red ((a) $(\tau,s)=(5,1)$) , blue  ((b) $(\tau,s)=(20,-1)$)
and brown  ((c) $(\tau,s)=(40,1)$).
The LO result from (a) (red) is better than the LO from (b) (blue), and is almost as good as the NLO (green)
to give the bound state energy, while the LO from (c) (brown) is a factor 3 larger than the exact value.
This seems counter intuitive, since the correlation function (c) contains low energy states more than other two as seen in Table~\ref{tab:param}. It might be that the low energy states are more affected by the presence of the bound state, which is not included in the correlation function.
The NNLO results (magenta), obtained by using all three correlation functions, almost reproduces the exact value of the bound state,
even though a bound state is not included in $R_L(x;\tau,s)$.

\section{Conclusion and discussion}
\label{sec:conclusion}
In this paper, we have demonstrated how the derivative expansion of the potential works to reproduce the scattering phase shift and the possible binding energy, by applying the HAL QCD method to highly non-local but solvable potential, the separable potential in quantum mechanics.
Our results strongly indicate that the derivative expansion in the HAL QCD potential method is NOT a way to reproduce the potential 
approximately. 
This can be easily seen from Fig.~ \ref{fig:Pot_c12}. 
In addition, the formal derivative expansion for the separable potential
does not give better results for the scattering phase shift, as we increase the order of the expansion.  
Instead, our results show that the derivative expansion in the HAL QCD method
gives an approximated way to extract scattering phase shifts from correlation functions as inputs.
More inputs we employ, better approximation for the phase shift we can obtain.
Even though the potential sometimes becomes singular in the coordinate space, as seen in Appendix A,  Fig.~\ref{fig:Pot_c12} at NNLO,
the scattering phase shift shows reasonable behavior and gives better approximation.
Interestingly, even without eigenfunction for the bound state, the binding energy is well reproduced in the HAL QCD method,
probably because a position of the  bound state in $k^2<0$ is well constrained by $k\cot\delta (k)$ at $k^2>0$
through analyticity. 

Lessons we obtain in this paper are as follows.
Even the LO approximations give reasonable results for the phase shift, and results can be improved as the order of the derivative expansion is increased.
Singular behaviors, which may appear at higher order such as the next-to-leading order in the derivative expansion,
are not obstructions for the potential method in principle.
In practice, however, less singular behavior is better to reduce statistical fluctuations.
Approximation for the scattering phase shift may break down at higher energy.
However, we know the applicable range of the method  in QCD,
since  the HAL QCD potential method, as well as the L\"uscher's finite volume method, work only below the inelastic threshold.  
In principle, by comparing results among different orders of the derivative expansion, we can estimate the size of systematics associated with the approximation. In practice, however, it is not so easy to extract  the potential at higher order reliably.
Therefore combining both the HAL QCD method and the finite volume method to extract scattering phase shifts
seems the best way to increase reliability and validity for lattice QCD calculations  on hadron interactions.
 
\section*{Acknowledgements}
We would like to thank Dr. Takumi Iritani for his contributions at the early stage of this work, 
Dr. Takumi Doi for his useful comments and suggestions, and
other members of the HAL QCD collaboration for useful discussions. 
SA is supported in part by the Grant-in-Aid of the Japanese Ministry of Education, Sciences and Technology, Sports and Culture (MEXT) for Scientific Research (Nos.~JP16H03978, JP18H05236),
by a priority issue (Elucidation of the fundamental laws and evolution of the universe) to be tackled by using Post ``K" Computer, and by Joint Institute for Computational Fundamental Science (JICFuS). 

\appendix
 \section{LO, NLO and NNLO potentials}
 \label{sec:appA}
 \begin{figure}[hbt]
  \begin{center}
\includegraphics[width=0.49\textwidth,clip]{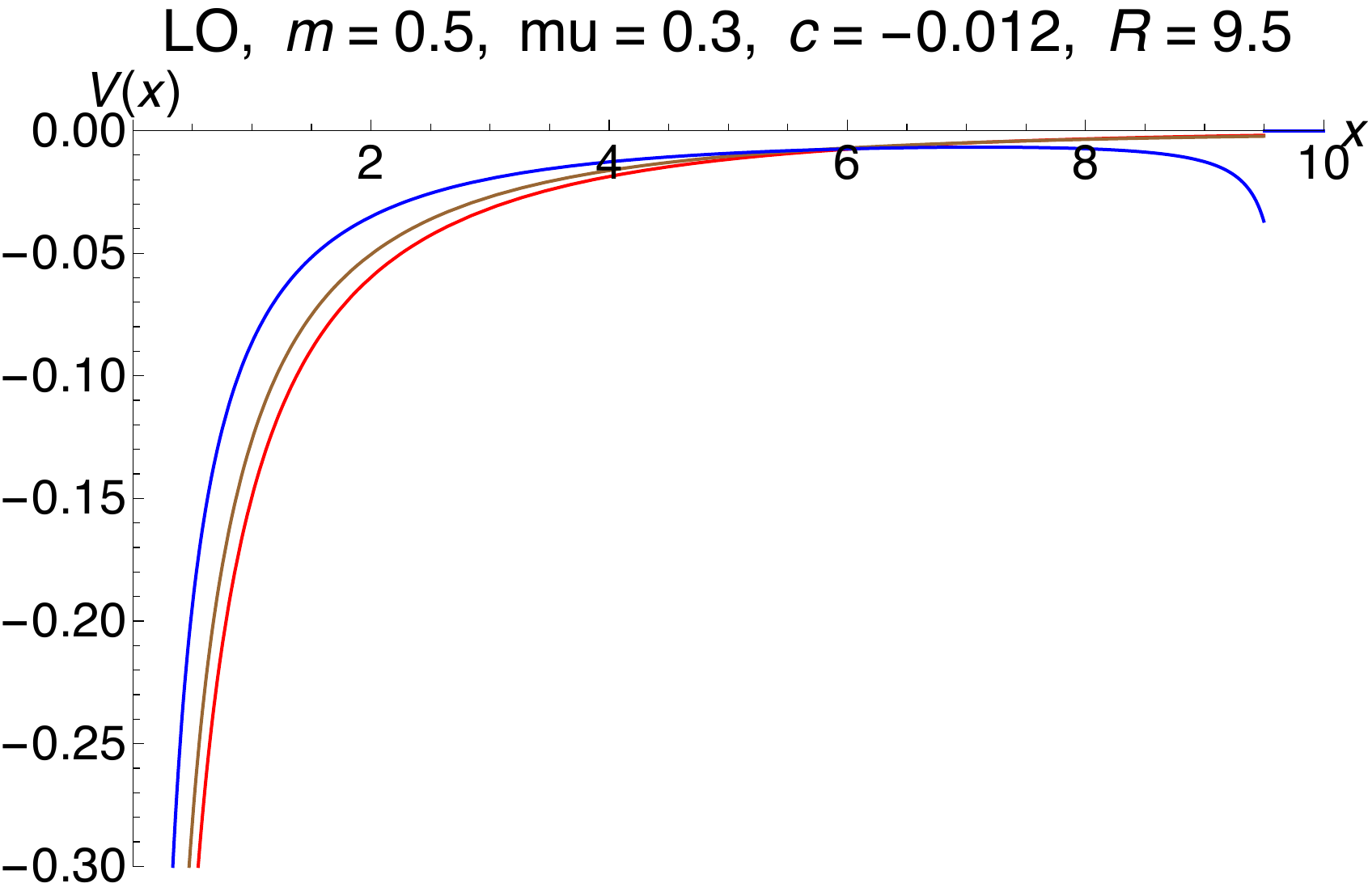}
 \includegraphics[width=0.49\textwidth,clip]{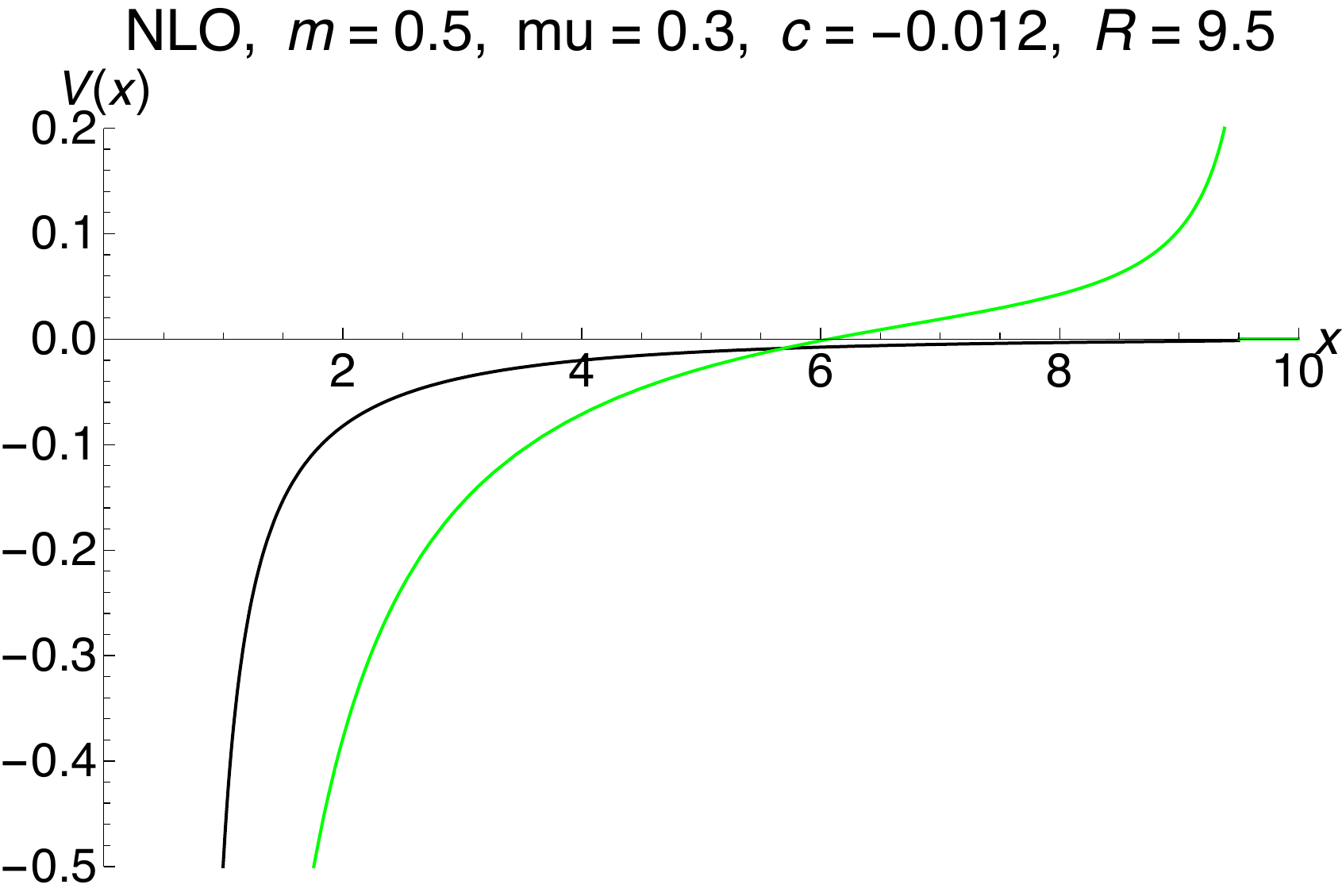}
  \includegraphics[width=0.5\textwidth,clip]{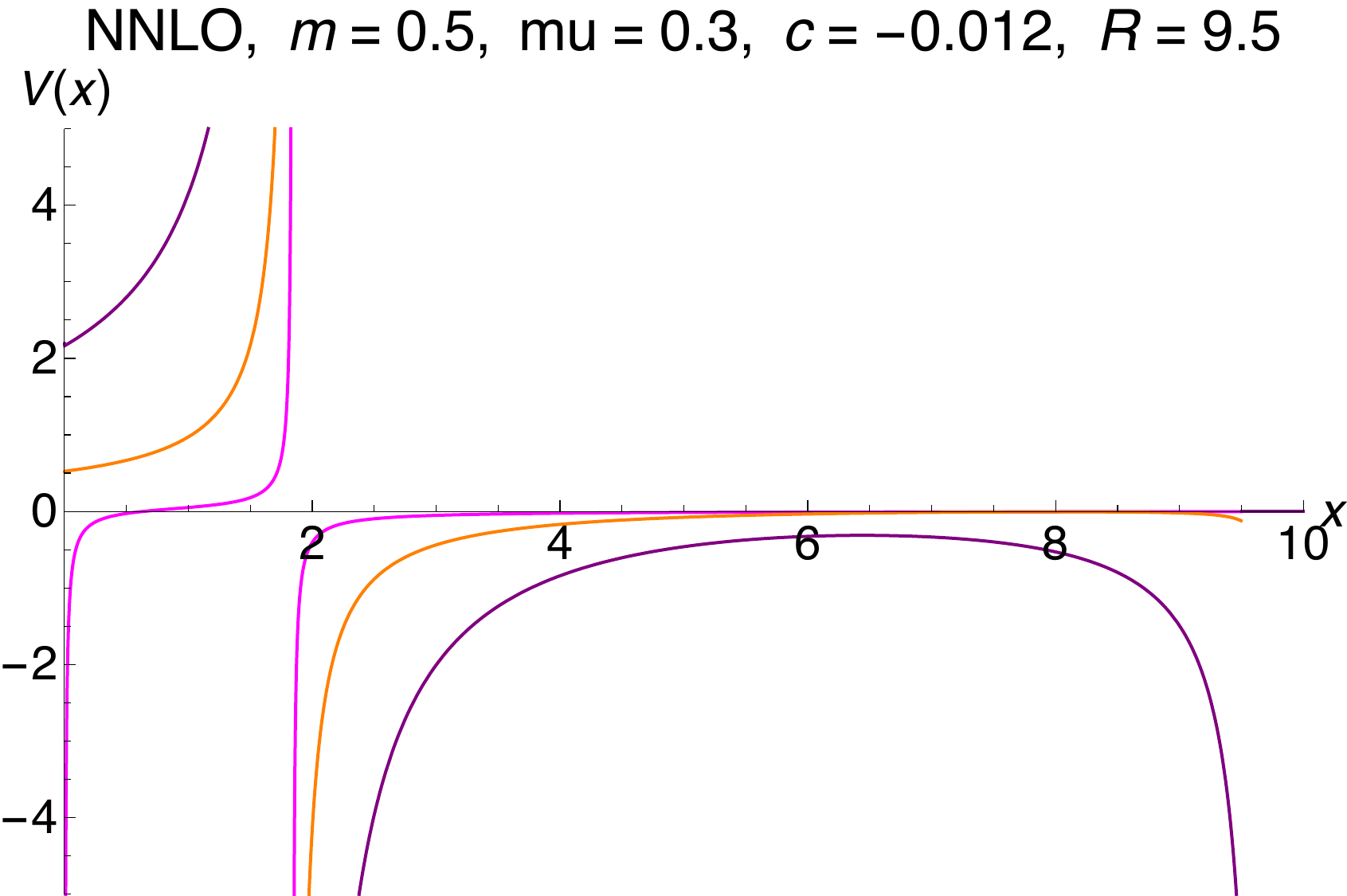}
  \end{center}
  \caption{
    \label{fig:Pot_c12}
  Shape of potentials at $m=0.5$,  $\mu=0.3$, $c=-0.012$ and $R=9.5$.
  (Upper left) The LO potential $V_{0,0} (x)$ from eigenfunctions at $k=0$ (red), $k=0.5\mu$ and $k=\mu$ (blue).
      (Upper-Right) The NLO potentials $V_{1,0}(x)$ (black) and $V_{1,1}(x)$ (green) from two eigenfunctions at   $k=0, \mu$.
      (Lower) The NNLO potentials $V_{2,0}(x)$ (magenta), $V_{2,1}(x)$ (orange) and $V_{2,2}(x)$ (purple) from all 3 eigenfunctions.  
        }
\end{figure}
As an representative example, we present shapes of the LO, NLO and NNLO potentials constructed form eigenfunctions
at $k=0, \mu/2, \mu$ with $m=0.5, \mu=0.3, c=-0.012$ and $R=9.5$.
In Fig.~\ref{fig:Pot_c12} (Upper-left),
we plot the LO potential $V_{0,0}(x)$, obtained from an eigenfunction at $k=0$ (red), $0.5\mu$ (brown) and $\mu$ (blue),
while we show coefficient functions of the NLO potentials $V_{1,0} (x)$ (black) and $V_{1,1}(x)$ (green), obtained from two eigenfunctions at $k=0,\mu$
in the upper-right of Fig.~\ref{fig:Pot_c12}.
The NNLO potential, obtained from all three eigenfunctions, is plotted in  Fig.~\ref{fig:Pot_c12} (Lower),
where $V_{2,0}(x)$ (magenta), $V_{2,1}(x)$ (orange) and $V_{2,2}(x)$ (purple) are shown.
As you see, the potential may become larger near the infrared cutoff, even for the LO potential.
In addition, each term of the NNLO potential show a very singular behavior at $x\simeq 2$.
As already observed in the main text, however, the corresponding phase shift  show a smooth behavior as a function of $k$
and improve an agreement with the exact result.

\section{Construction of correlation functions}
\label{sec:appB}
We define a correlation function through the Schr\"odinger equation as
\beqa
\left(-\partial_t -H_0\right) R(t,\vec x) &=&  \int d^3 y\, V(\vec x,\vec y) R(t,\vec y)
\eeqa
for $t\ge 0$ with an initial condition $R(0,\vec x) =\dfrac{\sigma^2 e^{-\sigma x}}{4\pi x}$,
 where $V(\vec x,\vec y)$ is the separable potential in the main text.

With this initial condition,  $R(t,\vec x)$ is given in \eqref{eq:Rcorr}, where
\beqa
f(\vec k) &=& \int d^3x\, \psi_k(\vec x)^\dagger R(0,\vec x), \quad
f_B =  \int d^3x\, \langle B \vert \vec x\rangle R(0,\vec x), 
\eeqa 
After straightforward but tedious calculations, we obtain
\beqa
R(t,\vec x) &=& R_0(t,\vec x) + R_1(t,\vec x),\\
K(t,\vec x) &:=& (-\partial_t -H_0) R(t,\vec x) =  (-\partial_t -H_0) R_1(t,\vec x) 
\eeqa
where
\beqa
R_0(t,\vec x) &=& {\sigma^2e^{ -\frac{m x^2}{2t}}\over 8\pi x}\left[
e^{\frac{t}{2m}(\sigma+\nu)^2}\left\{{\rm erf} \left(\sqrt{t\over 2m}(\sigma+\nu)\right)-1\right\}
\right. \nn
&-&\left. 
e^{\frac{t}{2m}(\sigma+\nu)^2}\left\{{\rm erf} \left(\sqrt{t\over 2m}(\sigma-\nu)\right)-1\right\}
\right]\\
&\longrightarrow& \left({m\over 2\pi t}\right)^{3\over 2} e^{ -\frac{m x^2}{2t}},
\quad \sigma\to\infty,
\eeqa
\beqa
R_1(t,\vec x)&=& {\sigma^2 \mu\gamma_B(\mu+\gamma_B)(e^{-\gamma_B x} -e^{-\mu x})\over
2\pi x (\sigma+\mu)(\sigma+\gamma_B)(\mu-\gamma_B)} e^{-E_B t} \theta(\gamma_B)
+{\sigma^2\mu(\mu+\gamma_B)^2\over 2\pi x (\sigma+\mu)}\nn
&\times&\left[ {\mu e^{{t\mu^2\over 2m}-\mu x}\over (\mu-\gamma_B)(3\mu+\gamma_B)(\sigma+\mu)}
\left\{ {\rm erf} \left(\sqrt{t\over 2m}(\nu-\mu)\right) +  {\rm erf}\left(\sqrt{t\over 2m}\mu\right)\right\}
\right. \nn
&-& {\gamma_B e^{t\gamma_B^2\over 2m}\over 2(\mu-\gamma_B)(\mu+\gamma_B)(\sigma+\gamma_B)}
\left\{ 
e^{-\gamma_B x}\left({\rm erf} \left(\sqrt{t\over 2m}(\nu-\gamma_B)\right)+\epsilon(\gamma_B) \right) 
 \right. \nn
&+&\left. e^{-\mu x} \left( {\rm erf}\left(\sqrt{t\over 2m}\gamma_B\right)-\epsilon(\gamma_B) \right)\right\}
-  {(2\mu+\gamma_B) e^{t(2\mu+\gamma_B)^2\over 2m}\over 2(3\mu+\gamma_B)(\mu+\gamma_B)(\sigma-2\mu-\gamma_B)}\nn
&\times&\left\{e^{(2\mu+\gamma_B) x}\left({\rm erf} \left(\sqrt{t\over 2m}(\nu+2\mu+\gamma_B)\right)-1 \right)
\right.\nn
&-&\left.e^{-\mu x}\left({\rm erf} \left(\sqrt{t\over 2m}(2\mu+\gamma_B)\right)-1 \right) 
\right\}
+ {\sigma e^{t\sigma^2\over 2m}\over (\sigma+\mu)(\sigma+\gamma_B)(\sigma-2\mu-\gamma_B)}\nn
&\times&\left. \left\{e^{\sigma x}\left({\rm erf} \left(\sqrt{t\over 2m}(\nu+\sigma)\right)-1 \right)
-e^{-\mu x}\left({\rm erf} \left(\sqrt{t\over 2m}\sigma\right)-1 \right) 
\right\} \right]\\
&\longrightarrow& { \mu\gamma_B(\mu+\gamma_B)(e^{-\gamma_B x} -e^{-\mu x})\over
2\pi x (\mu-\gamma_B)} e^{-E_B t} \theta(\gamma_B)
+{\mu(\mu+\gamma_B)^2\over 2\pi x}\nn
&\times&\left[ {\mu e^{{t\mu^2\over 2m}-\mu x}\over (\mu-\gamma_B)(3\mu+\gamma_B)}
\left\{ {\rm erf} \left(\sqrt{t\over 2m}(\nu-\mu)\right) +  {\rm erf}\left(\sqrt{t\over 2m}\mu\right)\right\}
\right. \nn
&-& {\gamma_B e^{t\gamma_B^2\over 2m}\over 2(\mu-\gamma_B)(\mu+\gamma_B)}
\left\{ 
e^{-\gamma_B x}\left({\rm erf} \left(\sqrt{t\over 2m}(\nu-\gamma_B)\right)+\epsilon(\gamma_B) \right) 
 \right. \nn
&+&\left. e^{-\mu x} \left( {\rm erf}\left(\sqrt{t\over 2m}\gamma_B\right)-\epsilon(\gamma_B) \right)\right\}
-  {(2\mu+\gamma_B) e^{t(2\mu+\gamma_B)^2\over 2m}\over 2(3\mu+\gamma_B)(\mu+\gamma_B)}\nn
&\times& \left\{e^{(2\mu+\gamma_B) x} \left({\rm erf} \left(\sqrt{t\over 2m}(\nu+2\mu+\gamma_B)\right)-1 \right)
\right. \nn
&-&\left.\left.
e^{-\mu x}\left({\rm erf} \left(\sqrt{t\over 2m}(2\mu+\gamma_B)\right)-1 \right) 
\right\}\right], \quad \sigma \to \infty,
\eeqa
\beqa
K(t,\vec x) &=&- {\sigma^2 \mu\gamma_B(\mu+\gamma_B)^2 e^{-\mu x}\over
4\pi m x (\sigma+\mu)(\sigma+\gamma_B)} e^{-E_B t} \theta(\gamma_B)
-{\sigma^2\mu(\mu+\gamma_B)^2\over 4\pi m x (\sigma+\mu)}e^{-\mu x} \nn
&\times&\left[ 
 {\gamma_B e^{t\gamma_B^2\over 2m}\over 2 (\sigma+\gamma_B)}
\left\{  {\rm erf}\left(\sqrt{t\over 2m}\gamma_B\right)-\epsilon(\gamma_B) \right\}\right.\nn
&+&  {(2\mu+\gamma_B) e^{t(2\mu+\gamma_B)^2\over 2m}\over 2(\sigma-2\mu-\gamma_B)}
\left\{{\rm erf}\left(\sqrt{t\over 2m}(2\mu+\gamma_B)\right)-1 \right\}\nn 
&-&\left. {\sigma (\sigma-\mu)e^{t\sigma^2\over 2m}\over (\sigma+\gamma_B)(\sigma-2\mu-\gamma_B)}
\left\{{\rm erf} \left(\sqrt{t\over 2m}\sigma\right)-1 \right\} \right] \\
&\longrightarrow& 
- {\mu\gamma_B(\mu+\gamma_B)^2 e^{-\mu x}\over
4\pi m x } e^{-E_B t} \theta(\gamma_B)
-{\mu(\mu+\gamma_B)^2\over 4\pi m x }e^{-\mu x} \nn
&\times&\left[ \sqrt{2m \over \pi t}+
 {\gamma_B e^{t\gamma_B^2\over 2m}\over 2}
\left\{  {\rm erf}\left(\sqrt{t\over 2m}\gamma_B\right)-\epsilon(\gamma_B) \right\}\right.\nn
&+&\left.  {(2\mu+\gamma_B) e^{t(2\mu+\gamma_B)^2\over 2m}\over 2}
\left\{{\rm erf}\left(\sqrt{t\over 2m}(2\mu+\gamma_B)\right)-1 \right\}\right],  \quad \sigma\to\infty.
\eeqa
 Here $\epsilon(x) :=2\theta(x)-1$, and the error function is defined as
 \beqa
 {\rm erf}(x) := {2\over \sqrt{\pi}}\int_0^x ds\, e^{-s^2}\to 1-{e^{-x^2}\over \sqrt{\pi} x}, \quad x\to\infty.
 \eeqa

 \bibliography{HALQCD}
\end{document}